\documentclass[11pt]{article} 
\pdfoutput=1

\usepackage{geometry} 
\usepackage{amsmath} 
\usepackage{amsfonts} 
\usepackage{amssymb}
\usepackage{graphicx} 
\usepackage[parfill]{parskip} 

\usepackage{setspace}

\usepackage{xspace}

\usepackage{lineno}  

\usepackage{natbib}
\usepackage{url}

\usepackage{caption}
\usepackage{subcaption}

\graphicspath{{:figs:}{figs/}}

\usepackage{booktabs} 
\usepackage{array} 
\usepackage{paralist} 
\usepackage{verbatim} 
\usepackage{subfig} 

\usepackage{fancyhdr} 
\usepackage{sectsty}
\allsectionsfont{\sffamily\mdseries\upshape} 

\usepackage[nottoc,notlof,notlot]{tocbibind} 
\usepackage[titles,subfigure]{tocloft} 

\usepackage{authblk}

\title{Toward an integrated system for fire, smoke, and air quality simulations}
\author[1]{Adam K. Kochanski}
\author[2]{Mary Ann Jenkins}
\author[3]{Kara Yedinak}
\author[4]{Jan Mandel}
\author[5]{Jonathan D. Beezley}
\author[6]{Brian Lamb}
\affil[1]{Department of Atmospheric Sciences, University of Utah, 135 S 1460 E, RM819, 84112 Salt Lake City, UT, USA, adam.kochanski@utah.edu}
\affil[3]{University of Idaho, College of Natural Resources, Forest, Rangeland,
and Fire Sciences Department, ID, USA}
\affil[4]{University of Colorado Denver, Denver, CO, USA}
\affil[5]{Kitware, Inc., 28 Corporate Drive, Clifton Park, New York 12065}
\affil[6]{Laboratory for Atmospheric Research, Department of Civil \& Environmental Engineering,
Washington State University, WA, USA}

\begin{document}
\maketitle


\begin{abstract}


In this study,  we describe how WRF-Sfire is coupled with WRF-Chem to construct WRFSC, 
an integrated forecast system for wildfire and smoke prediction.
The integrated forecast system has the advantage of not 
requiring a simple plume-rise model and assumptions about the size and heat 
release from the fire in order to determine fire emissions into the atmosphere.
With WRF-Sfire, wildfire spread, plume and plume-top heights are predicted directly, 
at every WRF time-step, providing comprehensive meteorology and fire emissions 
to the chemical transport model WRF-Chem.

Evaluation of WRFSC was based on comparisons between available observations to the results of two WRFSC simulations.

The study found overall good agreement between forecasted and observed
 fire spread and smoke transport for the Witch-Guejito fire.
 Also the simulated PM2.5 (fine particulate matter) peak concentrations matched the observations.
 However, the NO and ozone levels were underestimated in the simulations and the peak concentrations were mistimed.
Determining the terminal or plume-top height is one of the most important
aspects of simulating wildfire plume transport, and the study found overall good
agreement between simulated and observed plume-top heights, with some 
(10\% or less) underestimation by the simulations.
One of the most promising results of the study was the agreement between
passive-tracer modeled plume-top heights for the Barker Canyon fire simulation and observations.
This simulation took only 13h, with the first 24h forecast ready in almost 3h, making
it a possible operational tool for providing emission profiles for external chemical
transport models.  



\end{abstract}


\section{Introduction} \label{sec:introduction}

The United States 2009 Federal Wildland Fire Policy and the Clean Air Act have 
significantly broadened regulatory and management requirements by making 
assessment of air quality and visibility impacts from wildland fires mandatory.  
Because fire emissions can contribute to a violation of the National Ambient Air 
Quality Standards, wildland fire managers in the United States must consider 
the impacts of smoke on regional air quality and visibility.   
The adverse effects of smoke, described by \citet{hardy_etal:art:2001},
are particular concerns when planning prescribed burns or evacuations 
of populations at risk of smoke exposure.  
An increase in PM (Particulate Matter) levels due to forest fires may 
result in an increase of upper-respiratory track illness, asthma, and rhinitis,
serious health concerns associated with wildfire emissions \citep{emmanuel:art:2000}. 

Forecast tools of varying complexity are available to assess smoke dispersion at the regional scale. 
They range from simple Gaussian models such as VSMOKE \citep{lavdas:report:1996} and SASEM 
\citep{sestak_and_reibau:technote:1988} 
that predict the area affected by smoke based on fuel type, fire area, and wind conditions, to 
models such as CALPUFF \citep{scire:inproc:2000} for applications involving long-range transport, and 
finally to more complex multi-model systems like BlueSky \citep{larkin_etal:art:2009} that predict 
emissions, dispersion, and air quality effects associated with wildland fires. 

Even though plume models of various complexity exist, their physical descriptions of transport and 
dispersion of smoke from a wildland fire are generally inadequate.  
A performance evaluation of several widely used operational plume-rise models 
can be found in \citet{valmartin_etal:art:2012}, while a critical review of 
various smoke transport models can be found in \citet{goodrick_etal:art:2012}. 
Typically, assumptions are made about the size and heat release of the fire, and 
the vertical distribution of pollutants is represented either by a prescribed 
plume-rise and smoke-injection height or by simple smoke stack plume-rise algorithms. 
Also, the lack of physical representation of the spatial and temporal variability of the fire 
heat release in these models limits them in terms of realistic rendering of 
plume rise and smoke emissions. 



Clearly, the dispersion and transport of smoke from wildfires is a 
multi-disciplinary, multi-scale problem, affected directly by meteorology.   
Plume-rise is driven by the active fire area, total fire heat flux, and fuel moisture content of 
burning vegetation, all of which 
depend on local meteorology \citep{freitas_etal:art:2007, freitasetal:art:2010}.  
Atmospheric humidity and temperature affect fuel moisture, which in turn influences heat release
and fire spread.  
The wind field, atmospheric stability, and evolution of the atmospheric boundary layer 
affect pyro-plume development and smoke injection heights.   
Weather also governs smoke dispersion and the chemical reactions that 
take place as the smoke plume is advected downwind from the fire.  

Studies \citep[and references therein]{valmartin_etal:art:2012} 
indicate that accurate plume-rise prediction requires input of time-dependent 
meteorological conditions (winds, humidity, atmospheric stability,  
evolution of the atmospheric boundary layer), accurate sensible heat 
values from the burning area of a propagating wildfire, as well as 
the representation of the buoyancy and pressure gradient forces associated with pyro-convection. 
In this work we attempt to address these problems with an integrated smoke 
prediction system formed by coupling WRF-Sfire \citep{Mandel_etal:art:2011} with the chemical transport 
model WRF-Chem \citep{Grell-2005-FCO}. 
Unlike current modelling frameworks that prescribe fire activity and simplify fire 
emissions to a single plume whose vertical extent is estimated by a simple plume rise model,
the coupled WRF-Sfire-Chem aims to predict pyro-plume development, and
smoke dispersion and its air quality impacts, 
without separate, external model components, by comprehensively 
resolving fire spread, heat release during flaming combustion, fire 
emissions, fire plume rise, as well as downwind smoke dispersion and associated chemistry. 

Our concept is similar to that by \citet{trentmann_etal:art:2006},
who used the non-hydrostatic, high-resolution ATHAM (Active Tracer High Resolution 
Atmospheric Model) to simulate a forest fire plume. 
The major difference between \citeauthor{trentmann_etal:art:2006}'s approach and the one 
investigated here is that the heat and emission fluxes associated with the 
fire are predicted, not prescribed based on available observations.
Instead, the WRF-Sfire modelling system resolves explicitly 
fire progression and surface heat release depending on fuel and weather 
conditions forecasted by WRF.
Coupling WRF-Sfire with WRF-Chem enhances the capabilities of each,
providing numerical prediction of fire progression, plume rise, as well as short-range and 
long-range transport of pollution caused by wildfires.  

The paper is organized as follows.  
Descriptions of the WRF-Sfire, WRF-Chem, and the coupled WRF-Sfire/WRF-Chem model, 
hereafter refereed to as WRFSC, are given in Section \ref{sec:model_description}. 
The WRFSC simulation of the 2007 Witch-Guejito fire,
examined in Section \ref{sec:SA_simulations}, is used  
to evaluate how well WRFSC represents: the local dispersion  
(Section \ref{subsec:PM2.5_D04}) and the long-range dispersion 
(Section \ref{subsec:PM2.5_D03}) of fire PM2.5 emissions (particulate matter less than 2.5 $\mu$ in diameter); the 
impact of fire on air quality and the chemical transformation of 
fire emissions (Section \ref{subsec:MOZART}); and plume behaviour  
based on PM2.5 concentrations as smoke (Section \ref{sec:cofmass}).
The WRF-Sfire simulation of the 2012 Barker Canyon Fire,
described in Section \ref{sec:Barker_Fire}, is used to determine how well
smoke as a passive tracer represents --- at significantly less
computational cost compared to a WRFSC fire and smoke simulation 
with full chemistry ---  the plume-rise and smoke injection heights 
(Section \ref{sec:Barker_Evaluation}) that are currently
either prescribed or supplied by simple smoke stack plume-rise algorithms. 
The results of the study are summarized and concluded
in Section \ref{sec:Conclusions}.

\section{WRFSC model description} \label{sec:model_description}

\begin{figure}[t] 
\vspace*{2mm}
\begin{center}
\includegraphics[width=0.95\textwidth]{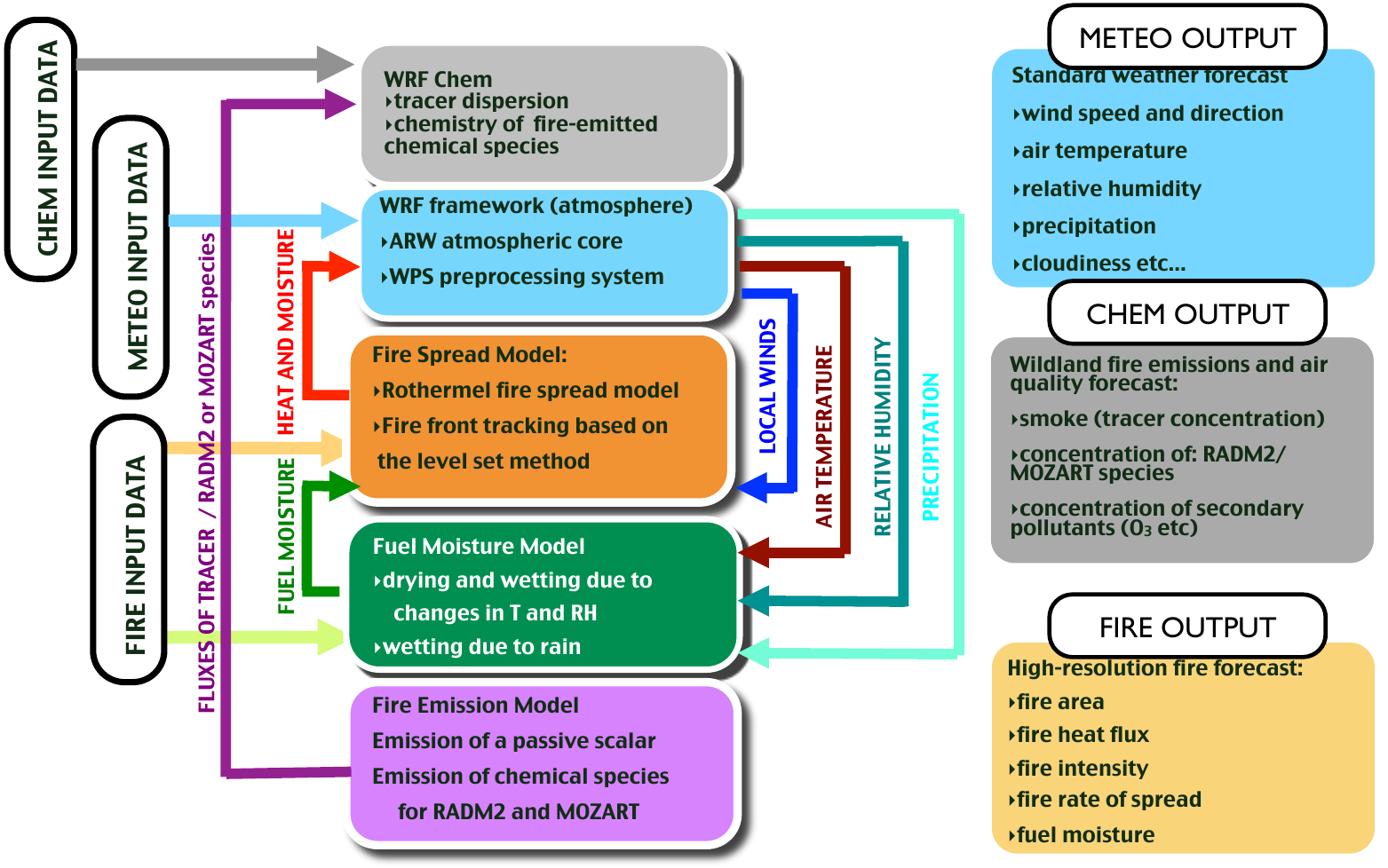} 
\end{center}
\caption{Diagram of WRF-Sfire coupled with fuel moisture model and 
WRF-Chem.\label{WSFC_diagram}}
\end{figure}

Fig.~\ref{WSFC_diagram} illustrates how each operating component in WRF-Sfire and
WRF-Chem couple to make up the integrated system WRFSC.
The core of WRFSC is the WRF-Sfire model \citep{Mandel_etal:art:2011}, 
a two-way coupled fire-atmosphere model based on WRF 
\citep{Skamarock_etal:technote:2008}.
WRF-Sfire's atmosphere-fire coupling estimates fire spread 
based on local meteorological conditions, taking into 
account the feedback between the fire and atmosphere \citep{clarkb:art:1996}.   
WRF-Sfire also contains a fuel moisture model that predicts fuel moisture based on local 
meteorology \citep{mandel_etal:NHESS:2014}.
WRF's nesting capabilities allow multi-scale domain 
configurations, where the outer domain, set at relatively low resolution, 
resolves the large-scale synoptic flow, while the gradually increasing 
resolution of inner domains provides representation of 
smaller and smaller scales required for realistic rendering of 
fire behaviour, fire convection, and smoke emissions.  
Coupling or feedback between inner and outer domains is two-way.  


To accommodate high-resolution fuel and elevation data, and provide sufficient 
accuracy for the fire spread computation without increasing computational cost, 
Sfire operates on a separate surface model fire grid, refined significantly with respect to 
WRF's innermost domain grid.
At every WRF time step, the near-surface wind from WRF is interpolated vertically to a 
logarithmic profile and horizontally to the surface fire grid to obtain height-specific wind 
that is input into the Rothermel model \citep{rothermel:techreport:1972} 
to determine the fireline's rate-of-spread.
A fuel type is selected from the 13 categories in \cite{anderson:techreport:1982}.
Each category includes a value for fuel mass, depth, density, surface-to-volume ratio, 
moisture of extinction, and mineral content.  
Based on these fuel properties and WRF-Sfire predicted winds, the surface fire-spread 
rate is computed at every refined mesh point.  
After ignition, the amount of fuel remaining is decreased exponentially over time, 
with the decay constant dependent on fuel properties following the BURNUP algorithm 
described in \citet{albini:techreport:1994}.  
The latent and sensible heat fluxes from the burning fuel are inserted into the 
lowest levels of WRF, assuming an exponential decay of the heat flux with height.  
A full description of the WRF-Sfire model can be found in Mandel et al. (2011). 
The current code and documentation are available from http://www.OpenWFM.org 
\citep{OpenWFM-2013-WUG}.  
A version from 2010 is distributed with the WRF release as WRF-Fire \citep{coen_etal:jam:2012,
OpenWFM-2012-FCW}.
 
\begin{figure*}[htbp] 
\vspace*{2mm}
\begin{center}
\includegraphics[width=0.95\textwidth]{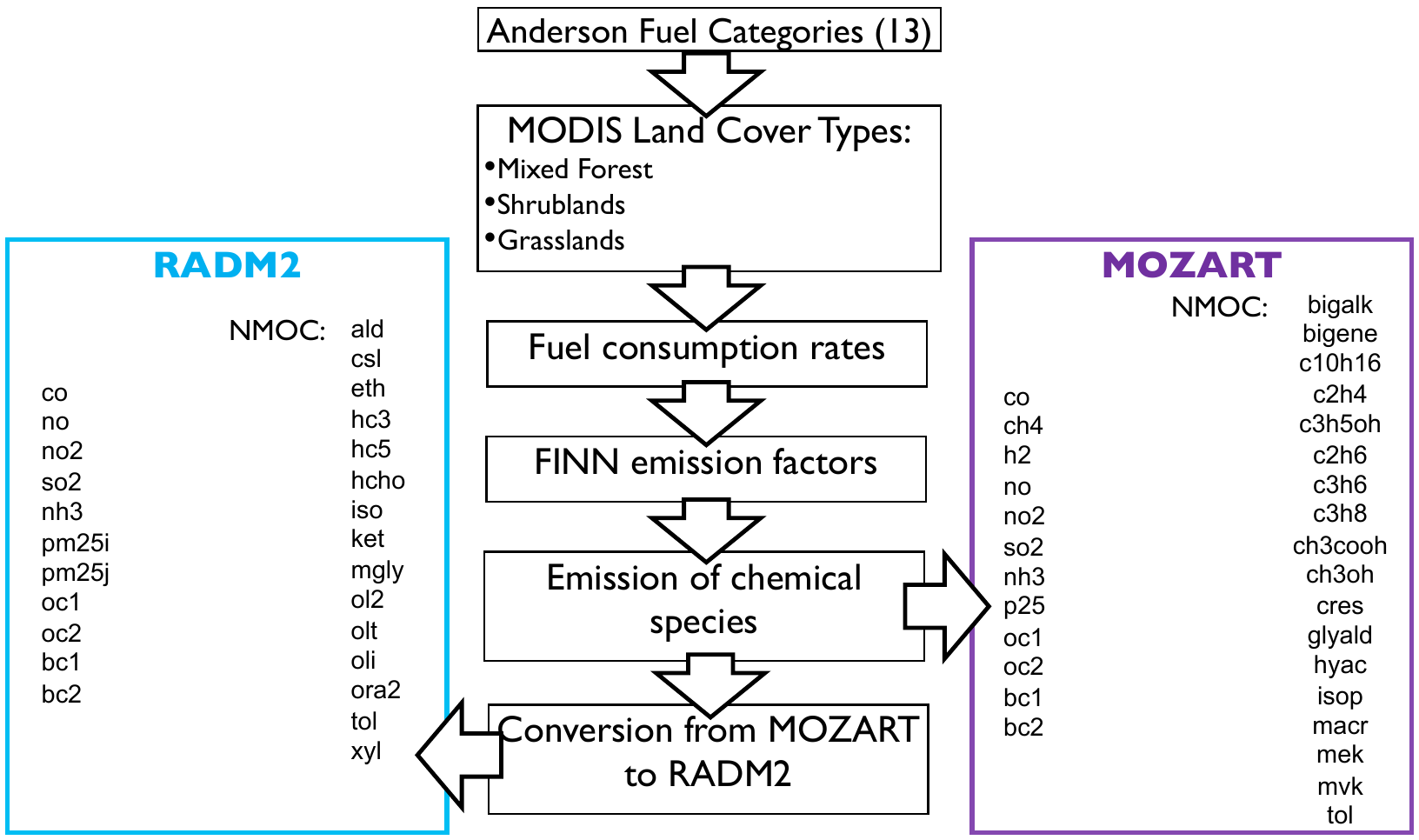}
\end{center}
\caption{Computation of emissions in WRFSC.}
\label{Fig_WSFC}
\end{figure*}

In the first step toward estimating smoke emissions, the Anderson fuel categories 
are converted into MODIS (Moderate Resolution Imaging Spectroradiometer)
Land Cover Types (see Fig.~\ref{Fig_WSFC}) 
compatible with FINN global emission inventory  \citep{wiedinmeyer_etal:art:2011}.   
After this conversion, combustion rates are computed for each fire-grid point based on the 
mass of fuel consumed within one WRF time step.
Once the fuel consumption is known, emission fluxes are computed as the 
products of the fuel-combustion rates and fuel-specific emission factors. 
The fire smoke emissions are then represented as a sum of fluxes of WRF-Chem-compatible chemical 
species listed in Fig.~\ref{Fig_WSFC}, which are ingested into the 
first model layer of WRF. 

Although WRFSC currently uses FINN global emission 
factors, the model accepts user-defined emission factors provided in one of the 
model configuration files. 
Since the FINN emission factors are compatible with the Model for Ozone and Related 
chemical Tracers (MOZART) \citep{emmons_etal:art:2010}, 
when the MOZART emission scheme is used, the chemical fluxes computed as described above are fed 
directly into the chemistry model.  
The RADM2 chemical scheme is also supported, through remapping between the MOZART
and the RADM2 chemical species \citep[their Table 7]{emmons_etal:art:2010}.    
Aside from the chemically reactive species, the FINN inventory allows  
estimates of emissions of particulate matter PM2.5 and PM10 which are integrated 
with the aerosol scheme as a part of WRF-Chem.

However, the treatment of the smoke as a mixture of chemically active species 
undergoing chemical and physical processes in the atmosphere is very computationally intensive.  
In addition to the complex mathematical representation of the chemical reactions,
each additional chemical species adds another three-dimensional scalar equation.
As a consequence, computational time for a WRFSC simulation with full chemistry is 
roughly 3.6 times longer than for a WRF-Sfire simulation. 
To avoid this computational cost, another smoke generation mechanism was developed, 
in which fire smoke is represented simply as a passive scalar. 
Instead of resolving the concentrations of all chemical species listed in 
Fig.~\ref{Fig_WSFC}, the scalar field of passive, chemically-inert tracers is added to WRF-Sfire 
computations. 
As with emissions fluxes for WRFSC, the surface flux of the tracer is computed based on 
the product of the combustion rate and the user-specified emission factor.
The execution of chemistry of fire-emitted chemical species in WRF-Chem is bypassed
(see Fig.~\ref{WSFC_diagram}),
and the WRF dynamical core handles transport of the passive smoke tracers directly.
The passive smoke tracers do not undergo any chemical reactions and do not interact 
radiatively or physically with the WRF environment. 
This simple smoke representation adds only about 20\% to the cost of 
running WRF-Sfire without a passive smoker tracer computation.

WRF-Chem contains a one-dimensional plume-rise module
\citep{freitas_etal:art:2007} to distribute fire emissions vertically as part of
WRF-Chem's preprocessing stage.  
The WRF-Chem plume-rise module releases about half of the fire emissions within the Atmospheric
Boundary Layer (ABL) and releases the other half into the atmosphere from injection heights above
the ABL \citep{pfisteretal:art:2011}. 
Using realistic injection heights for fire emissions is considered fundamental
to accurately simulating downwind transport and chemical evolution of fire plumes.
The smoke injection height is sometimes referred to as the terminal height,
based on the premise that the main detrainment (injection) mass layer
of plume convection is situated close to plume top.
With WRF-Sfire, however, plume-rise and plume-top heights are predicted 
directly, at every WRF time-step, with no need of a plume-rise module.

\section{WRFSC 2007 Witch-Guejito fire simulation} \label{sec:SA_simulations}

\subsection{Fire simulation setup} \label{sec:WSFC_WG_setup}

The Witch fire was first discovered on 21 October 2007 at 12:29 pm local 
time near State Highway 78 and Santa Ysabel 
(degrees latitude and longitude: 33.109168, -116.673794).  
The Guejito fire was first discovered on 22 October 2007 at 1:00 am local time in 
the Guejito Creek drainage, on the south side of State Highway 78 and 402 m west of 
Bandy Canyon Road (degrees latitude and longitude: 33.093694, -116.961639).  
A weather station at the Ramona Airport showed that, at that 
time, the ambient temperature was 22$^\circ$C, the relative humidity was 6\%, and 
15.6 m s$^{-1}$ winds were from the east-north-east, gusting up to 19.7 m s$^{-1}$ \citep{eidsmoe:report:2007}.
The cause of both fires was thought to be arcing between power lines.  
Both fires, progressing during strong, warm, and dry easterly Santa Ana winds,
burned a total 80,156 ha (801.56 km$^2$), making these the largest fires in 
California for 2007.  

\begin{figure*}[htbp] 
\vspace*{2mm}
\begin{center}
\includegraphics[width=1\textwidth]{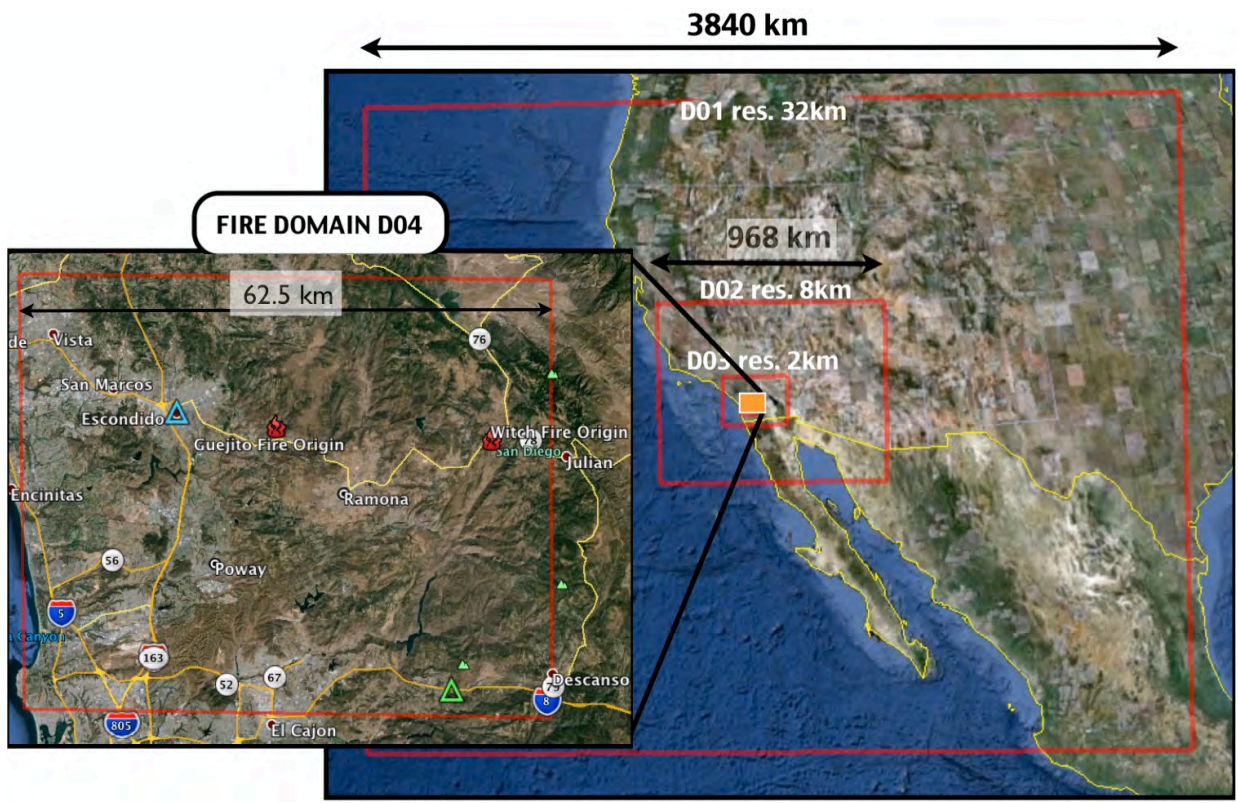}
\end{center}
\caption{The multi-scale WRF setup in this study, including locations of the 
Witch and Guejito fire origins and the air quality stations used for model 
validation (Escondido - blue triangle; Alpine - green triangle). 
Horizontal domain resolutions vary from 32km (D01) to 500m (D04). \label{WG-Domain-Config}}
\end{figure*}

The large-scale flow can interact with regional topography and land-use 
mosaic, creating specific local weather conditions that drive wildfire behaviour, 
which was the case in the Witch-Guejito fires.  
Therefore, in order to model development and movement of this large-scale weather 
system, including the Santa Ana winds it generated, together with the local 
circulation dictated by the complex topography of southern 
California, WRF was configured with four nested domains: D01, D02, D03, and D04, 
of horizontal-grid sizes 32km, 8km, 2km, and 500m, respectively. 
The model's vertically-stretch grid extended up to 15.4 km, 
with a surface layer roughly 20 m thick and the top-most vertical layer 
roughly 2000 m thick.  
The surface fire mesh located in domain D04 had a refinement 
ratio of 25, making the horizontal fire-grid cell size 20 m.  
Output from the fire simulation was saved every 10 minutes.  
The domain setup is shown in Fig.~\ref{WG-Domain-Config}. 
The fire model, Sfire, used 30m-resolution elevation and fuel data, 
while the atmospheric model, WRF, used approximately 1.5km-resolution MODIS 
land-use representation. 
Further details of this setup can be found in \citet{Kochanski-2013-RTS}.

The Witch-Guejito test case was initialized and driven by the NARR (North American 
Regional Reanalysis) data \citep{mesinger_etal:art:2006}, and run for a simulated time
period of 48h starting on 10.21.2007 5:00 am local time (12:00 UTC).  
The spatially-variable fuel moisture was initialized based on observations 
at the nearest weather station and then kept constant throughout the simulation.   
The observed relative humidity was extremely low
and, along with the dry and warm Santa Ana winds, inhibited 
fuel moisture recovery.

The Witch Fire was ignited in the Witch Creek area east of Ramona, California, 
almost 7h into the simulation. 
Both fires were initialized as point ignitions, and allowed to 
spread freely across the fuel mosaic driven by WRF-modelled local winds.   
The 48h simulation was done without updating the model 
state with current meteorological observations during the run. 
The WRF was provided with boundary conditions only at the initial 
stage and was not nudged towards the observations.  
This means that, unlike an actual operational forecast during which the WRF state 
is updated as the most current meteorological observations 
are assimilated, we did not interfere with WRFSC as it ran. 
A detailed description of model configuration and datasets used 
for model initialization, as well as a comparison between the 
simulated and observed fire progression, are found in \citet{Kochanski-2013-RTS}.

To isolate the impact of the simulated fire on the air quality,
no other chemical emissions were specified; 
i.e., WRFSC did not use the chemical boundary conditions and the standard idealized chemical 
profiles provided by WRF-Chem for chemical initialization.
The 48h WRFSC simulation with MOZART chemistry took 29h 56min on 324 CPUs of 
the University of Colorado Janus cluster (https://www2.cisl.ucar.edu/resources/janus). 

\subsection{PM2.5 dispersion inside fire domain D04} \label{subsec:PM2.5_D04}

Bulk particulate matter PM2.5 was used to 
represent the dispersion of pollutants emitted from the WRFSC simulated Witch-Guejito fires.
Not all the particulate matter in the simulation corresponded to 
that released from the Witch-Guejito fires. 
WRFSC-generated background PM2.5 concentration was present.
To distinguish between fire and non-fire PM2.5,
a PM2.5 threshold of 1 $\mu$g m$^{-3}$ was subtracted from the PM2.5 concentrations.
This threshold was found to be the lowest concentration 
that filtered out the background concentrations of PM2.5 without losing the 
fire signal from the freshly emitted PM2.5.

\begin{figure*}[htbp] 
\begin{center}
\includegraphics[width=0.95\textwidth]{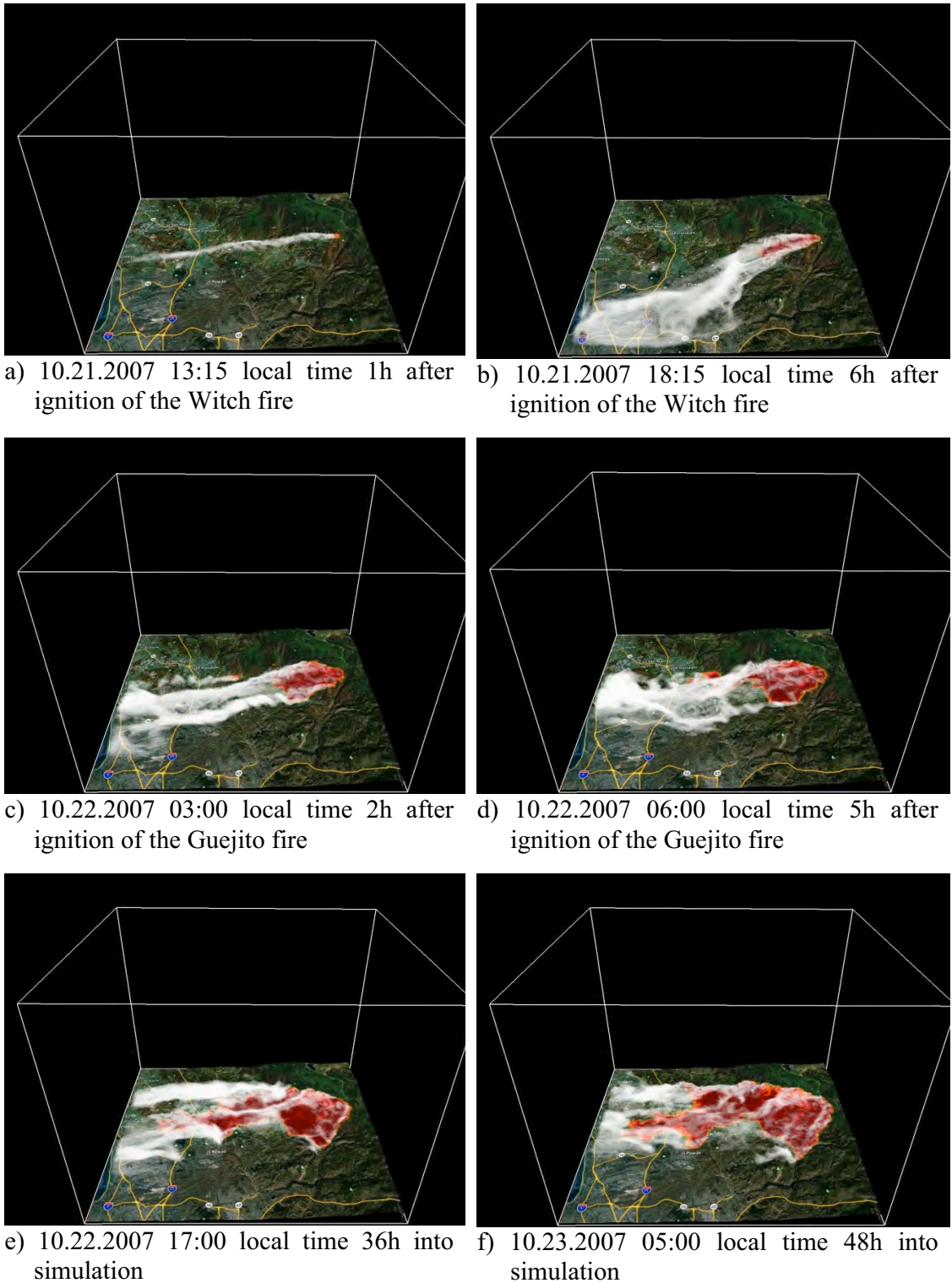}
\end{center}
\caption{Depiction of the Witch and Guejito fires at six different times in domain 
D04 of the simulation.  Smoke, shown in white, is represented PM2.5 
minus a background threshold PM2.5.   See text for further explanation.
Dark red fill represents burned-out fire area and orange fill shows active fire. 
\label{PM2.5_Smoke}}
\end{figure*}

Fig.~\ref{PM2.5_Smoke} shows the PM2.5 concentration results used to visualize 
smoke emissions from the Witch-Guejito fire.   
The Santa Ana wind blowing from east-north-east,
at a speed of up to 19 m s$^{-1}$ (68 k h$^{-1}$), with little variation in direction, 
rapidly pushed the Witch fire toward Encinitas and, 1 hour after ignition, 
confined the smoke to a long narrow downwind trajectory (Fig.~\ref{PM2.5_Smoke}a). 
Fig.~\ref{PM2.5_Smoke}b depicts smoke emissions 6 hours after ignition. 
Since WRFSC estimates emissions based on the combustion rate, emissions 
within the fire perimeter were not uniform.  
Smoke production was rapid and significant at the head of the fire and along 
the forward flanks, while inside the fire perimeter, where available fuel 
was depleted, emissions and smoke production were lower.  
As the fire progressed and its flanks expanded, the plume footprint also widened.    
The area affected by smoke was much larger than that in Fig.~\ref{PM2.5_Smoke}a.  
Fig.~\ref{PM2.5_Smoke}b shows how a slight change in wind direction in the western 
half of the fire domain pushed the smoke much further south compared to 
the previous result in Fig.~\ref{PM2.5_Smoke}a, 
with smoke covering a big part of the southwestern corner of fire domain.  
Note the sharp southeastern edge of the plume where steep canyons channeled the 
flow, limiting lateral smoke dispersion. 


Fig.~\ref{PM2.5_Smoke}c shows the fire perimeters and smoke propagation two 
hours after ignition of the Guejito fire.   
Smoke plumes from the two fires were separate, and even though the Guejito fire was much 
smaller, its contribution to the overall smoke pall is clearly visible in the 
southern part of Escondido. 
As both fires expanded and approached each other, their plumes 
combined (Fig.~\ref{PM2.5_Smoke}d) before the actual fire perimeters merged. 
Note the non-homogeneous smoke production at that moment.  
As the Witch fire approached Ramona (see Fig.~\ref{WG-Domain-Config} for location), 
its western edge became relatively inactive (Fig.~\ref{PM2.5_Smoke}e), 
which reduced smoke in that region. 
At the same time, there were two distinct smoke plumes, one formed 
by the Guejito fire and the northern flank of the Witch fire, and
the second formed primarily by the southwestern head of the Witch fire. 
Fig.~\ref{PM2.5_Smoke}f presents the situation at the end of the simulation 
after both fires merged.  
The former Guejito fire extended several kilometers southward and 
became a strong source of fire emissions affecting the Poway 
area (see Fig.~\ref{WG-Domain-Config} for location). 
Interior areas of active fire, as well as those along the eastern 
edge of the Witch-Guejito fire, contributed to the overall smoke pall.  
Fig.~\ref{PM2.5_Smoke}f shows a very complex fire and smoke emissions pattern. 
The wind continued to push the fire westward, with multiple hot 
spots all along the perimeter as well as within the interior.

How well the final perimeter of the 48h WRF-Sfire Witch-Guejito simulation compared
to the observed final fire perimeter is shown in \citet[their Fig.~8]{Kochanski-2013-RTS}.  
Although the agreement between observed and simulated final fire perimeters 
was relatively good overall, there were discrepancies, such as in the residental 
region north of Escondido, which was burnt in the WRF-Sfire simulation but not in reality.
(This is relevant to the results shown in Fig.~\ref{Escondido_AQ} in Section \ref{subsec:MOZART}.)

\begin{figure}
        \centering
        \begin{subfigure}[b]{0.54\textwidth}
                \includegraphics[width=\textwidth]{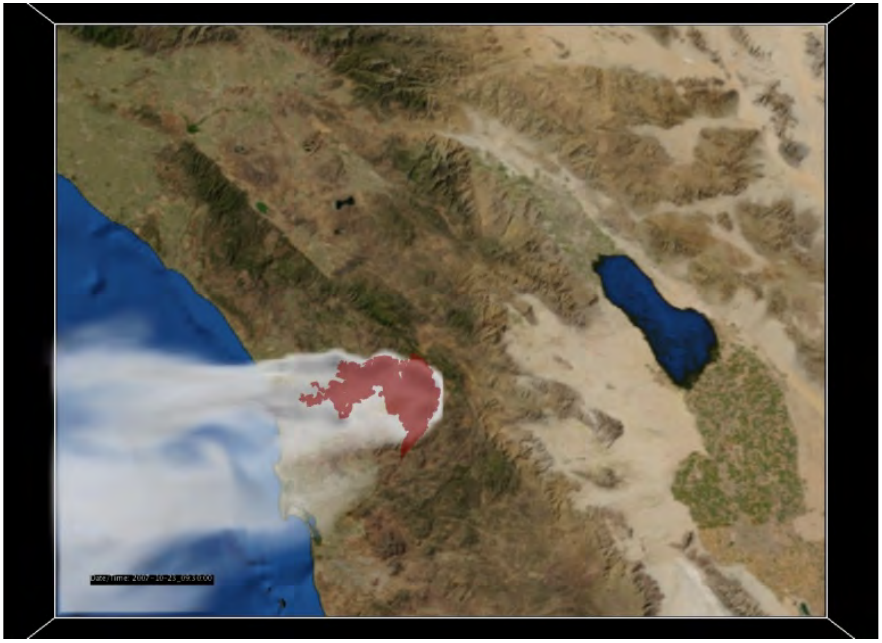}
                \caption{WRF domain D03 at time 10.23.2007 UTC}
                \label{fig:WRFvsMODISa}
        \end{subfigure}%
        ~ 
        \begin{subfigure}[b]{0.46\textwidth}
                \includegraphics[width=\textwidth]{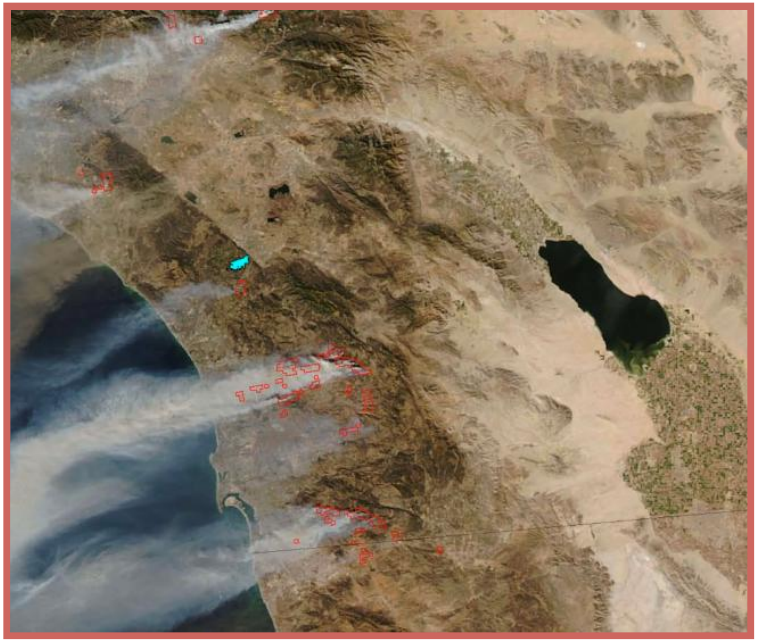}
                \caption{MODIS image at time 10.23.2007 UTC}
                \label{fig:WRFvsMODISb}
        \end{subfigure}
         \caption{Smoke dispersion within domain D03 (2km resolution) 
simulated by (a) WRFSC and (b) MODIS satellite image. 
The red color fill in (a) represents the fire area projected from the 
nested fire domain D04 (500m resolution). Red contours in (b) represent 
remotely detected hot spots (regions of the highest fire intensity).}\label{fig:WRFvsMODIS}
\end{figure}	 

\subsection{PM2.5 dispersion outside fire domain} \label{subsec:PM2.5_D03}

WRF's two-way coupling between multiple domains is used to transfer the 
PM2.5 smoke plume from the innermost, high-resolution fire domain to the coarser 
resolution outer domains to resolve large-scale smoke transport.  
An example of PM2.5 smoke dispersion within model domain D03 (2km resolution) is 
shown in Fig.~\ref{fig:WRFvsMODISa}.  
Fig.~\ref{fig:WRFvsMODISb} presents the corresponding satellite image from 
MODIS where estimates of active fire area obtained from MODIS fire radiative power
(FRP) thermal anomalies are compared to the modelled fire perimeter.
The MODIS image in Fig.~\ref{fig:WRFvsMODISb} includes smoke plumes produced from 
all fires in the region, while Fig.~\ref{fig:WRFvsMODISa} shows only the smoke from the 
WRFSC simulation of the Witch-Guejito fires.  
Nonetheless, there is at least a visible resemblance between the WRFSC 
simulated and MODIS-observed smoke dispersions.  
The aerial extent of each smoke plume is similar, and even 
some of the MODIS smoke dispersion 
features off the coast of San Diego are captured by WRFSC. 
The hot spots in the MODIS image in Fig.~\ref {fig:WRFvsMODISb} seem to be 
co-located along the simulated fire perimeter presented in Fig.~\ref{fig:WRFvsMODISa}.

\subsection{Fire impact on local air quality} \label{subsec:MOZART}
 
 \begin{figure}
         \centering
         \begin{subfigure}[b]{0.50\textwidth}
                 \includegraphics[width=\textwidth]{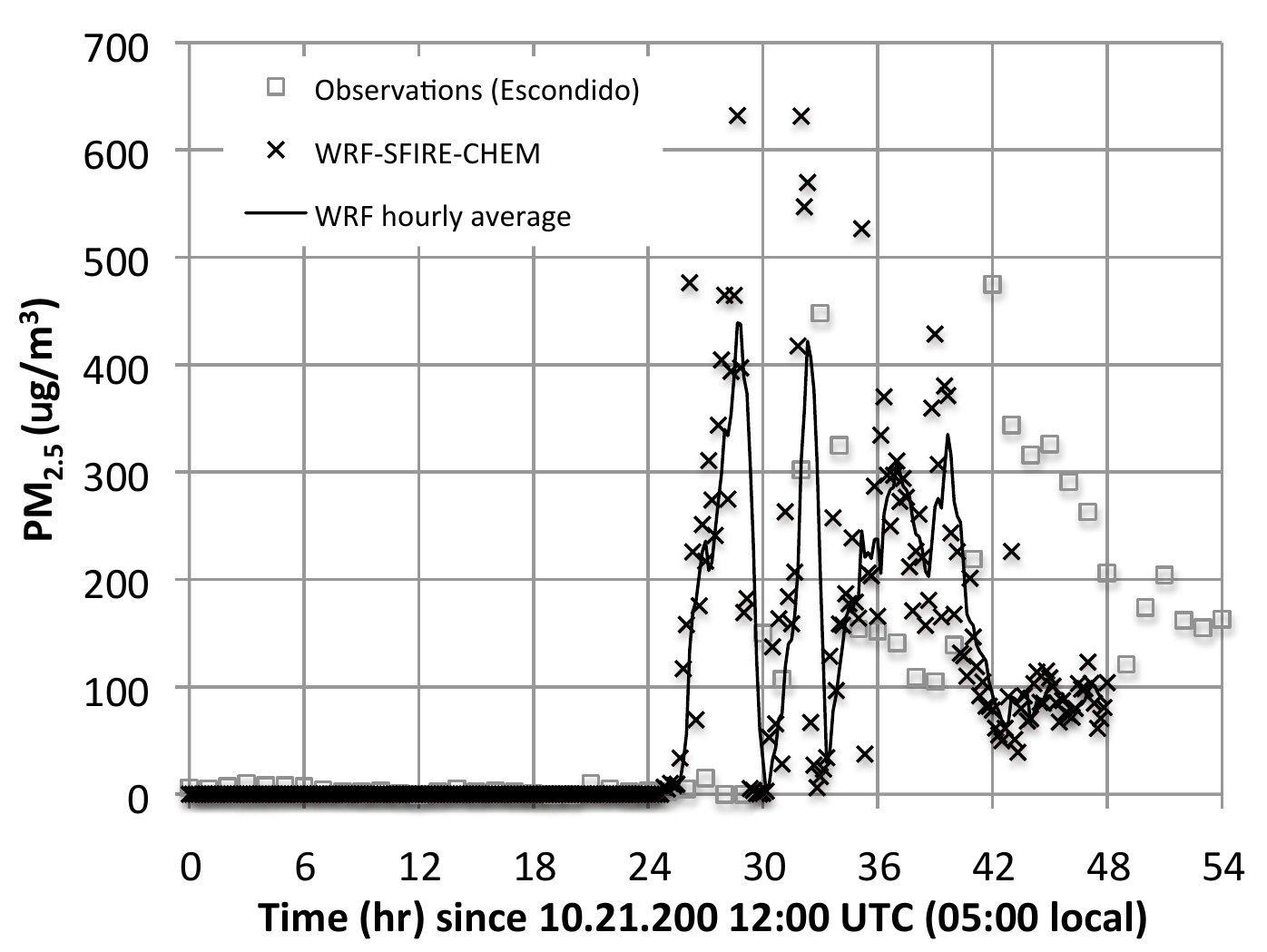}
                 \caption{}
                 \label{fig:Escondido_AQa}
         \end{subfigure} 
         \begin{subfigure}[b]{0.50\textwidth}
                 \includegraphics[width=\textwidth]{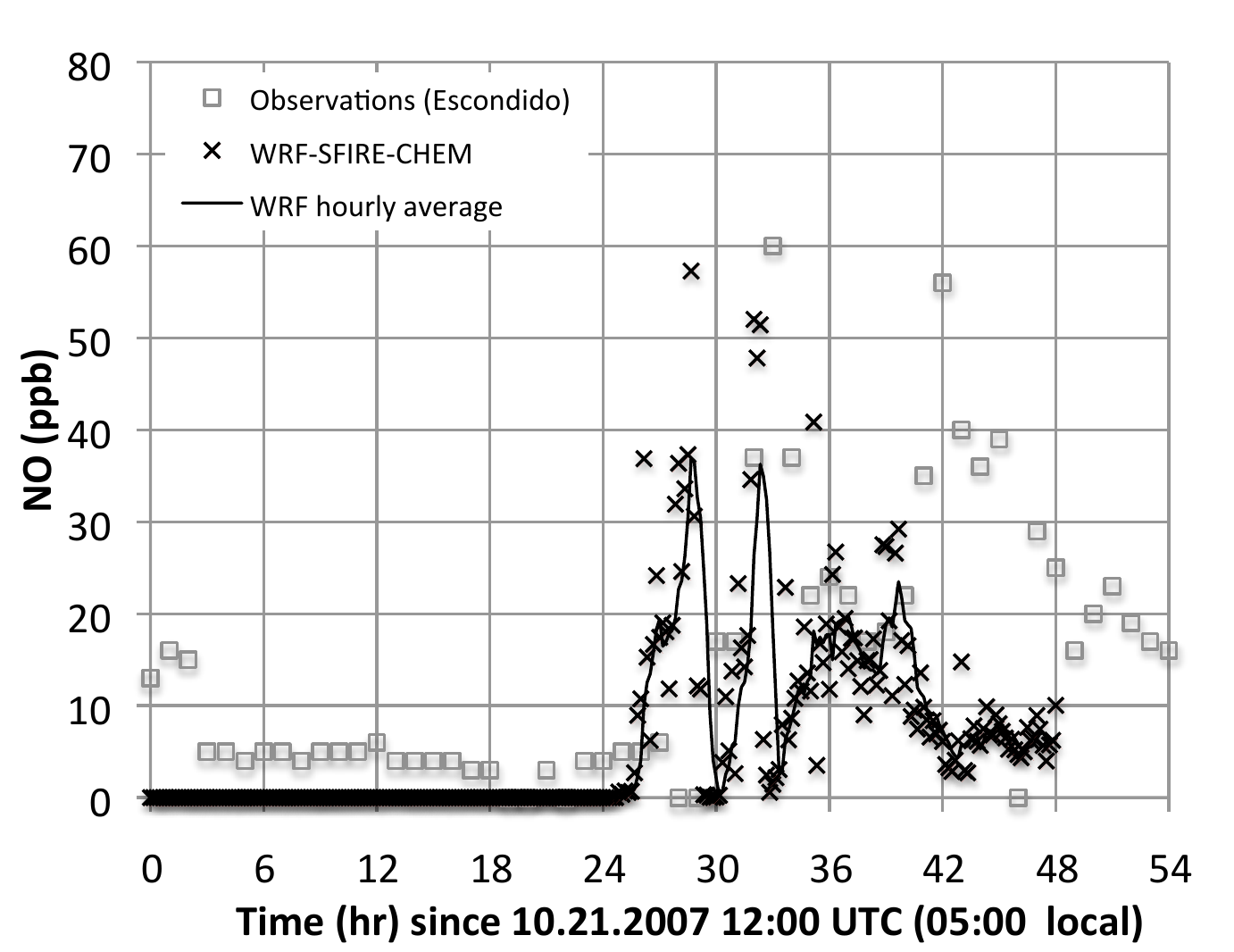}
                 \caption{}
                 \label{Escondido_AQb}
         \end{subfigure}
          \caption{Time series of the (a) PM2.5 ($\mu$g/m$^3$) and (b) NO (ppb) simulated by 
 WRFSC and observed by the Escondido air quality station.}
 \label{Escondido_AQ}
 \end{figure}	 

WRFSC's ability to represent concentrations of NO and fire-emitted fine particulate matter 
PM2.5 is shown in Fig.~\ref{Escondido_AQ}.  
The PM2.5 and NO fluxes are primary pollutants, emitted directly into the atmosphere 
from active fire grid points according to local fuel consumption rates 
and fuel-specific emission factors. 
Time series of 10-minute model output (marked with Xs) show PM2.5 concentrations greater than
observed and NO concentrations less than observed.
The 1h-running means of simulated PM2.5 concentrations for the Escondido station 
reached values close to those observed (450 versus 475$\mu$g m$^{-3}$),
while the 1h-running means of simulated NO concentrations were
significantly lower in magnitude compared to observations (38 ppb versus 60 ppb). 
WRFSC-predicted values peaked a couple of hours earlier than observed values.
The time shift between observed and simulated peaks can possibly be attributed to an 
overestimation in the north-west fire progression toward Escondido.
\citet{Kochanski-2013-RTS} determined that the 
simulated fire progressed approximately 10km further in the north-west
direction than the actual fire. 
 
In as complex a modelling system as WRFSC, it is not
possible to determine directly every reason for the discrepancies between the simulated 
and observed NO levels seen in Fig.~\ref{Escondido_AQ}b. 
One reason may be that the three MODIS Land Cover Types (i.e., mixed forest,
shrublands, and grasslands) are not detailed enough to represent the 
chemical smoke composition from fires spreading across 13 different Anderson fuel-bed 
categories. 
Another reason may be that WRF-Chem's global NO emission factors for grass and 
shrubs may be slightly lower than the actual NO emission factors for the fuel types in 
the Witch-Guejito region of Southern California.
Nonetheless, these differences between PM2.5 and NO simulated and 
observed are relatively low, and show promise for improved fire-emissions forecasts.
  
\begin{figure}
         \centering
         \begin{subfigure}[b]{0.50\textwidth}
                 \includegraphics[width=\textwidth]{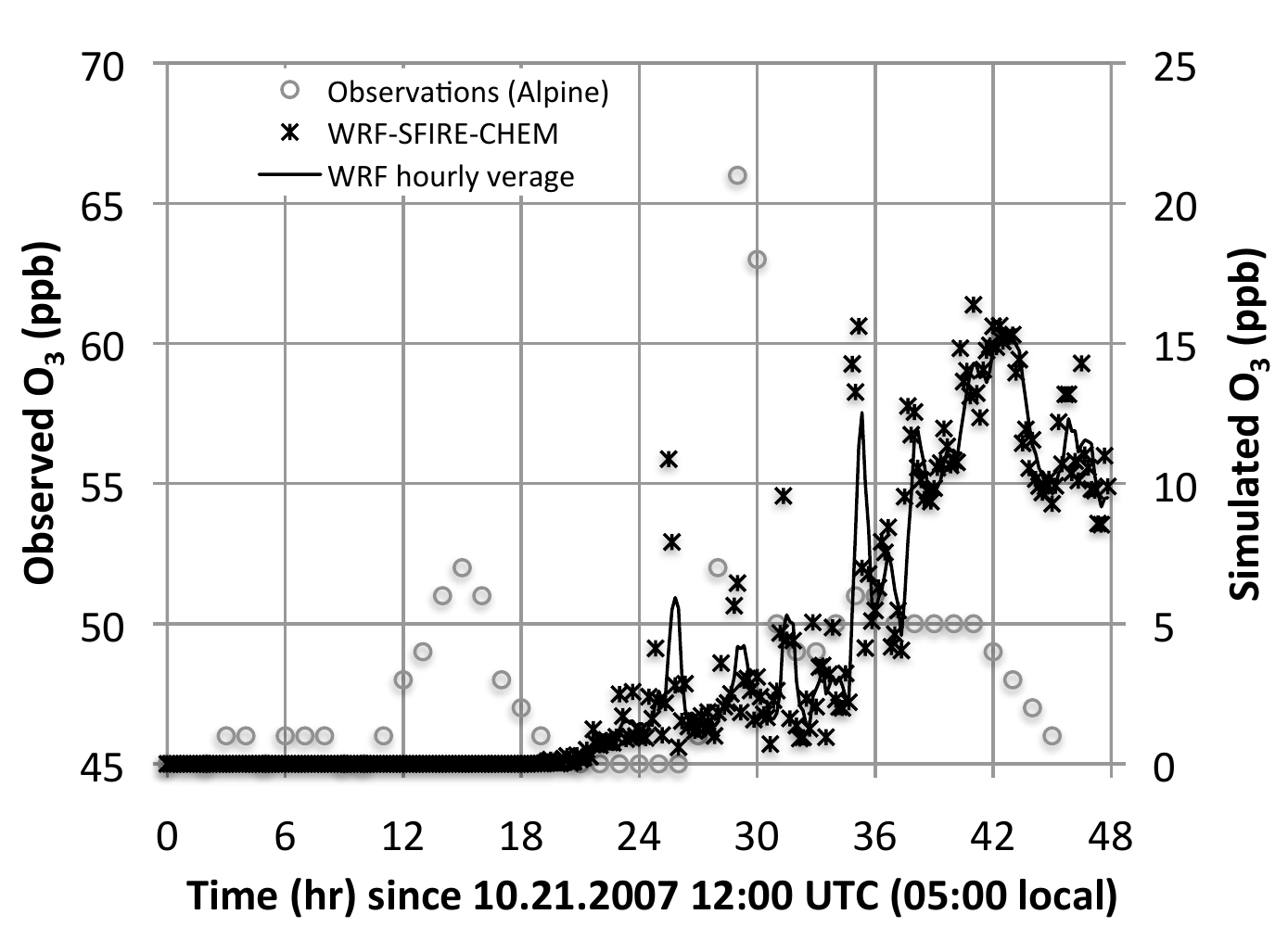}
                 \caption{}
                 \label{fig:Fig_7a}
         \end{subfigure}
         \begin{subfigure}[b]{0.50\textwidth}
                 \includegraphics[width=\textwidth]{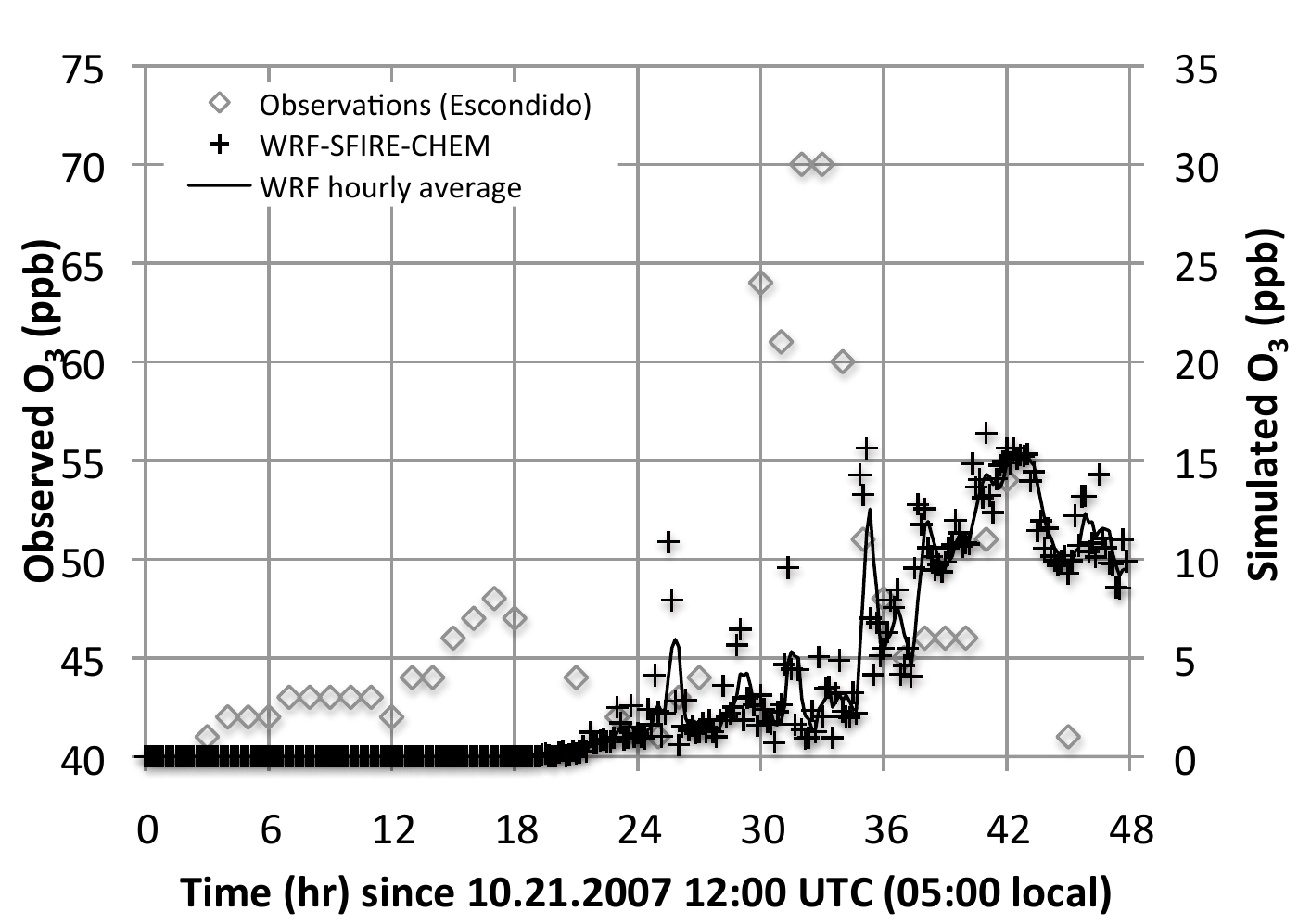}
                 \caption{}
                 \label{fig:Fig_7b}
         \end{subfigure}
         \caption{Time series of WRFSC simulated and of observed ozone concentrations at air
quality stations (a) Alpine and (b) Escondido. Plus signs (+) denote 10-minute WRFSC output
values, solid line denotes 1h running-means of WRFSC 10-minute output values.} \label{A-E-ozone}
 \end{figure}

Integration of WRF-Sfire with the WRF-Chem chemistry modules should also 
capture the chemical plume transformation in the atmosphere, that then 
represents the effect of smoke on secondary pollutants, such as ozone. 
A test of WRFSC's ability to represent ozone concentrations is presented in                   
Fig.~\ref{A-E-ozone}. The locations of
Escondido and Alpine air quality stations used in this study
are shown in Fig.~\ref{WG-Domain-Config} (blue and green triangles).
 
Surface ozone measurements are affected by other emissions, such as traffic, 
biogenic, and emissions from other fires, that were not included in the WRFSC simulation.
As described in Section \ref{sec:model_description}, the chemical component of WRFSC 
was set up in a simplified way, neglecting all anthropogenic and biogenic emissions, 
except those from the simulated fire, and 
the WRFSC fire simulation was started with zero ozone. 
Therefore, when comparing the modelled to surface-station ozone observations 
shown in Fig.~\ref{A-E-ozone}, the focus is on differences in amplitudes, 
rather than absolute values, of the fire-related peaks in ozone concentrations.
The time series of surface ozone concentrations in Fig.~\ref{A-E-ozone} are plotted using 
dual vertical axes. 
For easier comparison between the observed and simulated ozone peaks, 
both axes are scaled similarly, but the axes for observed ozone concentrations start 
at pre-fire values of 45 and 40 ppb, while the axis origin for the simulated results 
starts at 0. 
The pre-fire values of 45 and 40 ppb in Fig.~\ref{A-E-ozone} 
match well the pre-fire ozone concentrations 
range of 40.1 to 44.5 ppb estimated by \cite{bytnerowicz_etal:inproc:2010}.
 
Fig.~\ref{A-E-ozone}a shows that the simulated fire-caused peak in ozone 
concentration at the Alpine station was lower in magnitude than the observed one. 
 Observations indicate a 21 ppb increase in the ozone concentration, while 
 simulated ozone increased by 16 ppb. 
 Fig.~\ref{A-E-ozone}b shows that the underestimation of ozone concentration by 
 WRFSC is even more evident at the Escondido station. 
 Measurements indicate an increase of about 27 ppb while modelled ozone concentration 
 peaked at 16 ppb. 
 The exact reason for these discrepancies is hard to identify.
 It is possible that the underestimation in the fire emissions of precursor NO 
 (Fig.~\ref{Escondido_AQ}) contributed to this problem. 
 Simulated ozone peaks are also delayed compared to the observations. 
 Station measurements show ozone concentrations peaking late in the morning in 
 Alpine and early afternoon in Escondido, while modelled ozone concentrations peak 
 later in the evening. 
 
We lack the observational data to determine if the fire emission inventory 
is represented correctly, making further analysis of the chemical mechanisms in the 
chemistry component of WRFSC beyond the scope to this work.  
It is possible that the discrepancies between simulated and observed results 
shown Fig.~\ref{Escondido_AQ} may be partly due to 
the simplified chemical setup used in this study and that the simulation period 
was too short for the chemical model to spin up. 
Also,  WRF-Sfire accounts for flaming emissions only, and smoldering emissions are
neglected. 
 
\subsection{Plume-top heights based on PM2.5 concentrations}\label{sec:cofmass}

\begin{figure*}[htbp] 
 \vspace*{2mm}
 \begin{center}
 \includegraphics[width=1\textwidth]{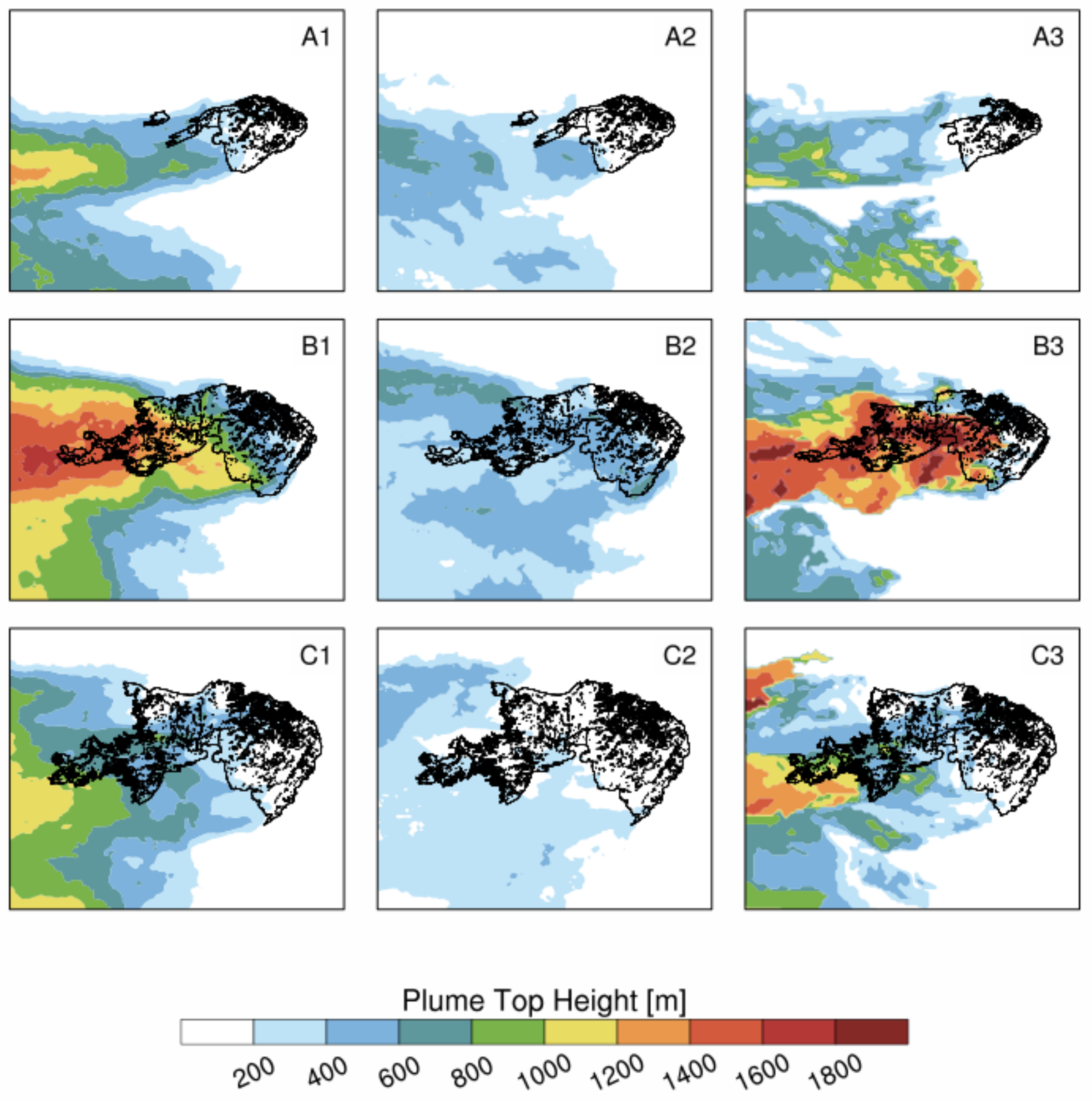}
 \end{center}
 \caption{The rows depict three time frames during the WRFSC Witch-Guejito
 simulation:   
 (A) night of the 21st; (B) daytime of the 22nd; and (C) night of the 22nd.
 The columns, from left to right, represent the: (1) 6-hr averaged plume-top heights; 
 (2) standard deviations of 6-hr averaged plume-top heights; and (3) a snapshot of 
 plume-top height 3 hours into each 6-hr time frame.  
 The burned area for each time period is outlined in black.
 The plots cover domain D04 (62.5 x 52.5 km).\label{Fig_Kara_plume_hts}}
 \end{figure*}

AGL (Above Ground Level) plume-top heights based on PM2.5 were calculated for 
each 10-minute model output and grid column in D04 (62.5 km x 52.5 km) with fire 
PM2.5 present.
The plume-top heights were then averaged over 6-h time intervals representing 
night (10:00pm to 4:00am local time) and day (10:00am to 4:00pm local time) time periods.
These 6-h averages are shown in the left column of Fig.~\ref{Fig_Kara_plume_hts}, where row A  
are the night plume-top heights of 21 October,
row B are the day plume-top heights of 22 October, and row C
are the night plume-top heights of 22 October. 
The middle column gives the corresponding standard deviations for 
these same time periods.
The right column shows a single snapshot of the 
plume-top heights at 3 hours into each 6-hr averaged time period.  
The corresponding burned area for each time period is shown in black.
 
The left and right columns in Fig.~\ref{Fig_Kara_plume_hts} shows strong spatial
and temporal variability in plume-top heights 
with maximum values between 1100 to 2200 m AGL downwind of the fire and
increasing in magnitude as the fire grew in size.
Maximum plume-top heights were:
1329 m (top left); 1709 (middle left); 1159 (bottom left);
1291 m (top right); 2163 (middle right); 1898 (bottom right).
The middle column shows that the standard deviations of the time-averaged 
heights were often as large as 600 m, also indicating considerable variation over 
each 6-hr period from particular portions of the fire perimeter.   
Fig.~\ref{Fig_Kara_plume_hts} illustrates clearly the non-Gaussian 
horizontal spread of fire plume.
Plots, like those in the right column, but for the entire simulation
(not shown), captured multiple plume-rise peaks, associated  
with multiple plume updraft cores, a phenomenon known to occur often in 
wildland fires \citep{goodrick_etal:art:2012}.
Based on these results, it may not be accurate to represent 
the plume and fire emissions of a propagating wildfire in terms of a 
single Gaussian-shaped plume and terminal or plume-ejection height.
 
\begin{figure*}[htbp]
\vspace*{2mm}
\begin{center}
\includegraphics[width=0.7\textwidth]{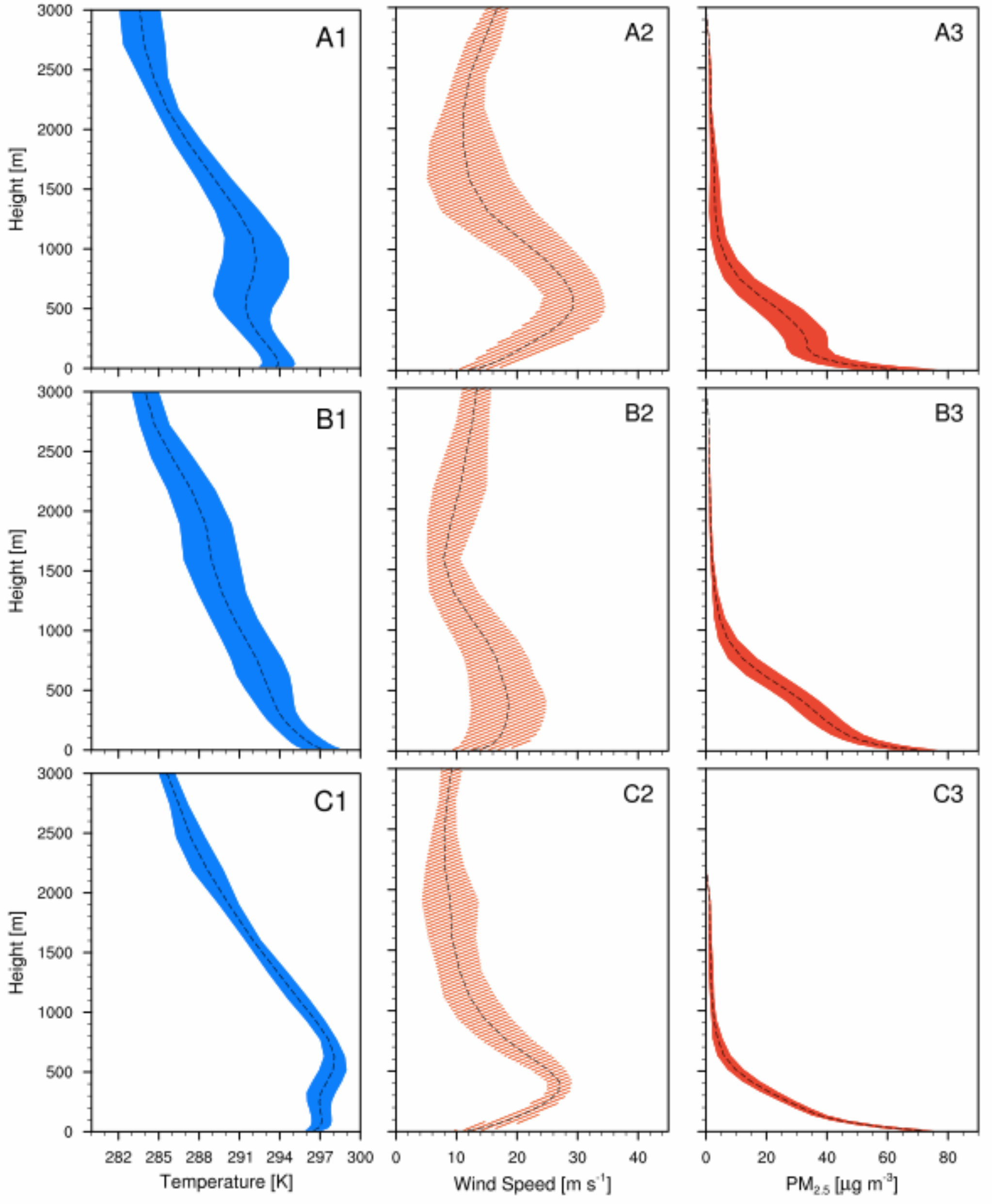}
\end{center}
\caption{
Vertical profiles of 6-hr averages of temperature (left column), wind speed (middle column), 
and total PM2.5 concentrations (right column) for night of the 21st (top row), 
daytime of the 22nd (middle row), and night of the 22nd (bottom row). 
PM2.5 concentrations $\le$ 1.0 $\mu$g m$^{-3}$
were omitted from PM2.5 averages, and the height is AGL.  See text for further details.
\label{fig:T_WS_diurnal}}
\end{figure*}

Plume-rise models, such as \citet{freitas_etal:art:2007, freitasetal:art:2010},  
used to simulate injection heights in CTMs (Chemical Transport Models),
require as input temperature and wind-speed profiles in order to determine
the vertical distribution of fire emissions in the atmosphere near the
fire source (i.e., injection heights).
Fig.~\ref{fig:T_WS_diurnal} shows the temperature, wind-speed, and fire emission profiles 
for the WRFSC Witch-Guejito fire.
Vertical profiles of the near-fire 6-hr averaged temperature (left column, blue) and
wind speed (middle column, light red), as well as domain-averaged 
PM2.5 (right column, red) for the (A) night of the 21st, (B) daytime of the
22nd, and (C) night of the 22nd are given.
Close to the fire means locations upwind of the burn area, and 
therefore not influenced by the burn, with a displacement of 
roughly one horizontal WRF grid cell away from the burn area.  

Vertical profiles of daytime 6-hr averaged ambient temperatures 
(Fig.~\ref{fig:T_WS_diurnal} B1) show elevated surface temperatures  
compared to the nighttime surface temperatures on the first, pre- Santa Ana day (Fig.~\ref{fig:T_WS_diurnal} A1), 
but very similar to the nighttime temperatures during the Santa Ana event
(Fig.~\ref{fig:T_WS_diurnal} C1).    
The temperature profiles also show daytime-nighttime variation in both height 
and strength of an inversion layer topping the atmospheric boundary layer.
The 6-hr averaged vertical wind speed profiles (Fig.~\ref{fig:T_WS_diurnal} B-C 2) 
show a surface jet below 500 m AGL, descending toward the surface, and causing 
strongest overall wind speed magnitude and variability during the daytime  
(Fig.~\ref{fig:T_WS_diurnal} B2). 
Both the nighttime and daytime vertical profiles for PM2.5 (Fig.~\ref{fig:T_WS_diurnal}, right panels)
shows surface PM2.5 concentrations exceeding the
daily EPA standards of 35 $\mu$g m$^{-3}$.

The smoke injection height, sometimes known as the terminal 
height, is fundamental to accurately representing fire
emissions in CTMs \citep{valmartin_etal:art:2012},
and the vertical distribution of fire emissions (if available) 
is used to locate the smoke injection height.
Fig.~\ref{fig:T_WS_diurnal}, column 3, 
does not however, indicate a single, simple terminal plume height.  
The PM2.5 profiles indicate instead, depending on the PM2.5 concentration
cutoff or decay rate used, a range plume-ejection heights between 1000 
to 1500 m AGL.
And this range matches the range of plume-top heights found in
Fig.~\ref{Fig_Kara_plume_hts}.
Based on the PM2.5 profiles, 
the minimum plume injection height is assumed to be
approximately 1000 m AGL.

 

The MISR (Multi-angle Imaging SpectroRadiometer) satellite data retrieval 
during the time period of the Witch-Guejito WRFSC simulation was 
also used to evaluate the WRFSC's ability to represent plume-top heights.  
There was no MISR retrieval for the Witch-Guejito fire.  
Plume height estimates based on available MISR data satellite retrievals of two wildland
fires in the vicinity of the Witch-Guejito fire during the time period of the 
simulation are given in Table \ref{Table:MISR-W-G}.
A straightforward averaging of the MISR plume heights in Table \ref{Table:MISR-W-G} 
gives an average median plume height of 814 m and an average standard 
deviation 292 m, and an average plume-top height of 1236 m and an average standard
deviation of 379 m.  
These results compare well to range of the simulated plume-top 
heights (1000 to 1500 m AGL) suggested by Figs.~\ref{Fig_Kara_plume_hts} 
and \ref{fig:T_WS_diurnal}.

 \begin{table}
 \vspace*{2mm}
 \centering
 \begin{tabular}{|c|c|c|c|c|c|c|c|} \hline
 Longitude & Latitude &Plume     &Plume   &Median  &Top      &SDev  &FRP  \\ 
           &          &Perimeter &Area    &Plume   &Plume    &      &     \\
           &          &Length    &        &Height  &Height   &      &     \\
           &          &[m]       &[km$^2$]&[m AGL] &[m AGL]  &[m]   &MW   \\
 \hline\hline
 -117.441  &  33.244  & 70       &259    &943     &1260      &269   &NA  \\  
 -116.562  &  32.623  & 269      &2168   &684     &1017      &314  &1061  \\  
 \hline
 \end{tabular}
 \caption{MISR plume retrieval data for wildfires in the vicinity of Witch-Guejito fire 
on date 2007-10-21 and time 18:39:52 UTC. SDev = Standard Derivation. 
FRP = Fire Radiative power.  MW = Megawatts.  See text for details.}
 \label{Table:MISR-W-G}
 \vspace*{2mm}
 \end{table}

 
\begin{figure*}[htbp]
\vspace*{2mm}
\begin{center}
\includegraphics[width=0.80\textwidth]{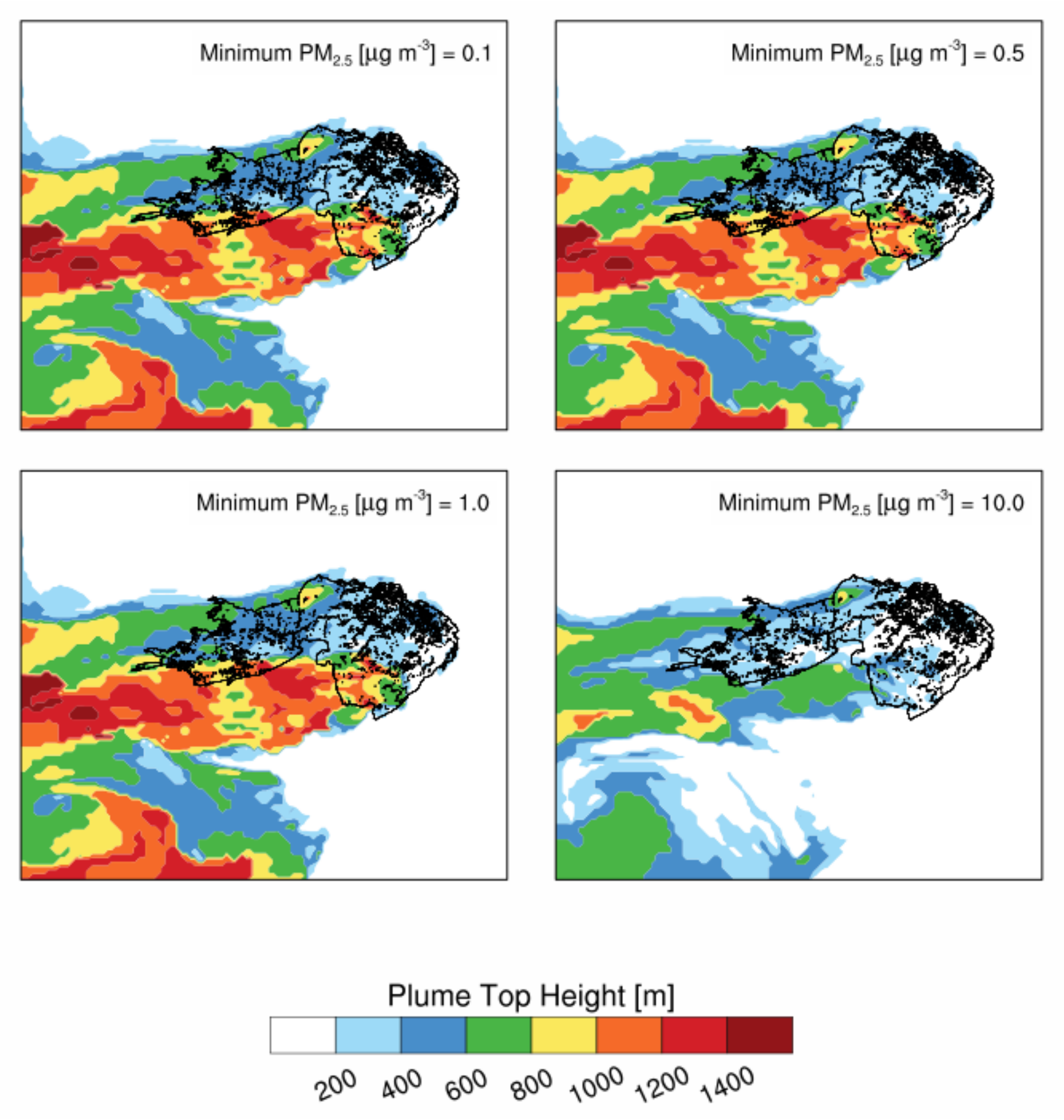}
\end{center}
\caption{ 
Contours show the 6-h averaged maximum plume top heights for
the Witch-Guijito WRFSC simulation based on 
four minimum PM2.5 filters:
upper left 0.1 $\mu$g m$^{-3}$; 
upper right 0.5 $\mu$g m$^{-3}$;
bottom left 1.0 $\mu$g m$^{-3}$;
bottom right 10.0 $\mu$g m$^{-3}$.
\label{PM2.5-filter_test}}
\end{figure*}

Fig.~\ref{PM2.5-filter_test} is similar to Fig.~\ref{Fig_Kara_plume_hts}
except that it illustrates the impact of a threshold background level PM2.5 
on plume-top height calculations.
Different filter quantities were tested, and the plume-top heights
shown in Fig.~\ref{PM2.5-filter_test} indicate that a PM2.5 filter in a range from 0.1 - 10 
$\mu$g m$^{-3}$ assure robust plume-top heights estimates insensitive to the particular value of the threshold. 
Maximum plume top height was 1574 m for all PM2.5 filters $\le$ 1.0 $\mu$g m$^{-3}$,
but 1092 m, almost 500 m lower, for PM2.5 filter 10.0 $\mu$g m$^{-3}$.
 
\section{WRF-Sfire 2012 Barker Canyon fire simulation}  \label{sec:Barker_Fire}
 
\subsection{Fire simulation setup} \label{sec:WRF-Sfire_BC_setup}

The high computational costs and the discrepancies between observed and 
modelled ozone estimates for the Witch-Guejto WRFSC simulation
indicate that WRFSC fire and smoke simulations with full chemistry are not 
yet practical.
When only plume-top height estimates are needed and  
chemistry and smoke dispersion are handled by an external chemistry model,
running WRFSC, the full coupled version of WRF-Sfire-Chem, is not necessary.
Here we present the results of a WRF-Sfire simulation of the 2012 Barker 
Canyon fire, where fire smoke is represented by passive (non-reactive) tracers, 
as opposed to a mixture of chemically reactive species.

\begin{figure*}[htbp]
 \vspace*{2mm}
 \begin{center}
 \includegraphics[width=1\textwidth]{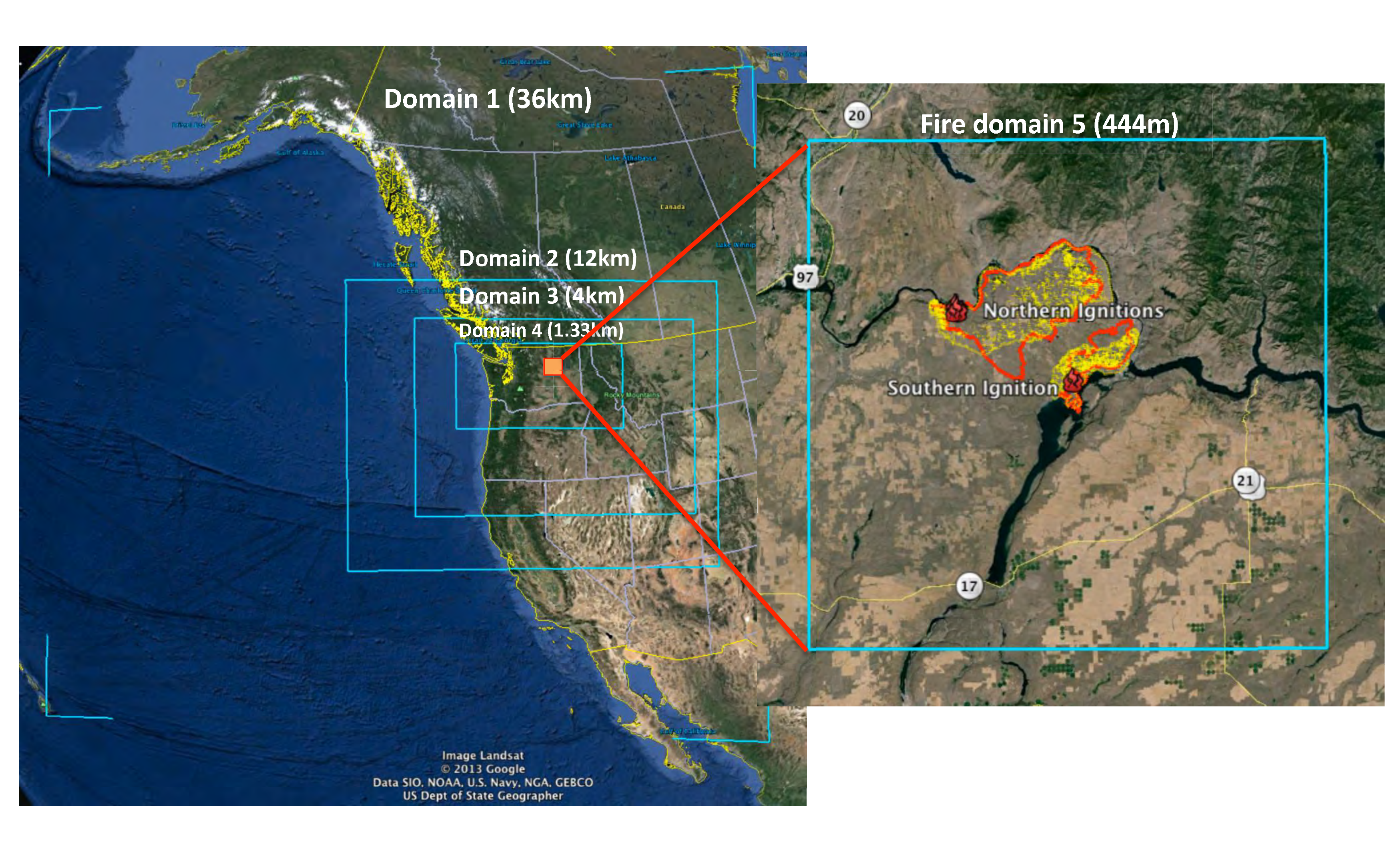}
 \end{center}
 \caption{Domain setup used for Barker Canyon fire simulation. Red contours represent the observed 
 final fire perimeters; yellow shading represents the WRF-Sfire simulated final perimeters. The
Barker Canyon fire started approximately on 2012-09-13 00:45 PDT. \label{Barker_fire_domain}}
 \end{figure*}

The Barker Canyon fire was ignited on 09.08.2012 by a series of lightning strikes 
in its northern branch around 19:20 and its southern branch around 20:00 local time
(Figure \ref{Barker_fire_domain}). 
The two fires within the complex grew dramatically during a wind event over September 10-11 
burning a total area of 32,842 ha (328.42 km$^2$) by September 19. 
 
The Barker Canyon fire simulation used 5 nested domains of resolutions 
from 36km to 444m (i.e., 1/3 grid ratio) (Fig.~\ref{Barker_fire_domain} and
Table \ref{Table:MISR-BCF}).
The fire model mesh refinement was set to 20 for the innermost fire domain 
(i.e., 22.2m resolution). 
The NARR reanalysis supplied boundary conditions to WRF throughout the simulation.
High resolution topography and fuel data (at 30m resolution), available from Landfire 
(http://www.landfire.gov), were used by Sfire, the fire component of the system. 
Initial fuel moisture in the 1h class was computed following \citet{wagner_pickett:techreport:1985}
using initial WRF-simulated air temperature and relative humidity fields. 
The initial 10h, 100h and 1000h fuel moistures were set to 4\% , 8\% , and 7\%, respectively, 
based on the national fuel moisture maps. 
The simulation was run with the fuel moisture model \citep{mandel_etal:NHESS:2014} 
to account for the significant fuel moisture changes observed for 
the Barker Canyon fire.

\begin{center}
    \begin{tabular}{ | p{1.5cm} | p{1.8cm} |p{2cm} |p{1.7cm} | p{0.8cm} |p{1.1cm}
|p{2.5cm} |p{1.6cm} |}
    \hline
     Domain Number & Number of grid points (x y z) & Atmospheric model
resolution & Fire model resolution  & Time step & Surface Model & PBL scheme
& Cumulus scheme \\ \hline
     D01 & 151x127x37 & 36km & - & 180s & Noah & Mellor-Yamada-Janjic &
Kain-Fritsch \\ \hline
     D02 & 182x142x37 & 12km & - & 90s  & Noah &  Mellor-Yamada-Janjic &
Kain-Fritsch\\ \hline
     D03 & 406x283x37 & 4km & - & 30s & Noah &  Mellor-Yamada-Janjic  &
Kain-Fritsch\\ \hline
     D04 & 712x364x37 &1.33km & - & 10s & Noah &  Mellor-Yamada-Janjic & -\\
\hline
     D05 & 196x193x37 & 444m & 22.2m & 3.3s  & Noah &  Mellor-Yamada-Janjic
& -\\ 
        \hline
    \end{tabular}
\label{Table:MISR-BCF}
\end{center}

The southern branch of the fire started from the ignition point reported 
by the Incident Information System (http://inciweb.nwcg.gov). 
The northern branch was ignited using locations of four lightning strikes observed 
within the fire perimeter on the day of ignition. 
The approximate locations of the fire ignition points 
are presented in Fig.~\ref{Barker_fire_domain}.
The Barker Canyon fire simulation was started on 09.09.2013 at 
00 UTC (09.08.2012 17:00 PDT) 
and then run for 96h without observational nudging.
The 96h simulation took 12h 52min on 640 CPUs, with the first 24h forecast ready in 
3h 13min.
 
\subsection{Barker Canyon fire with smoke as a passive tracer} \label{sec:Barker_Evaluation}
 
 \begin{figure*}[htbp]
 \vspace*{2mm}
 \begin{center}
 \includegraphics[width=0.7\textwidth]{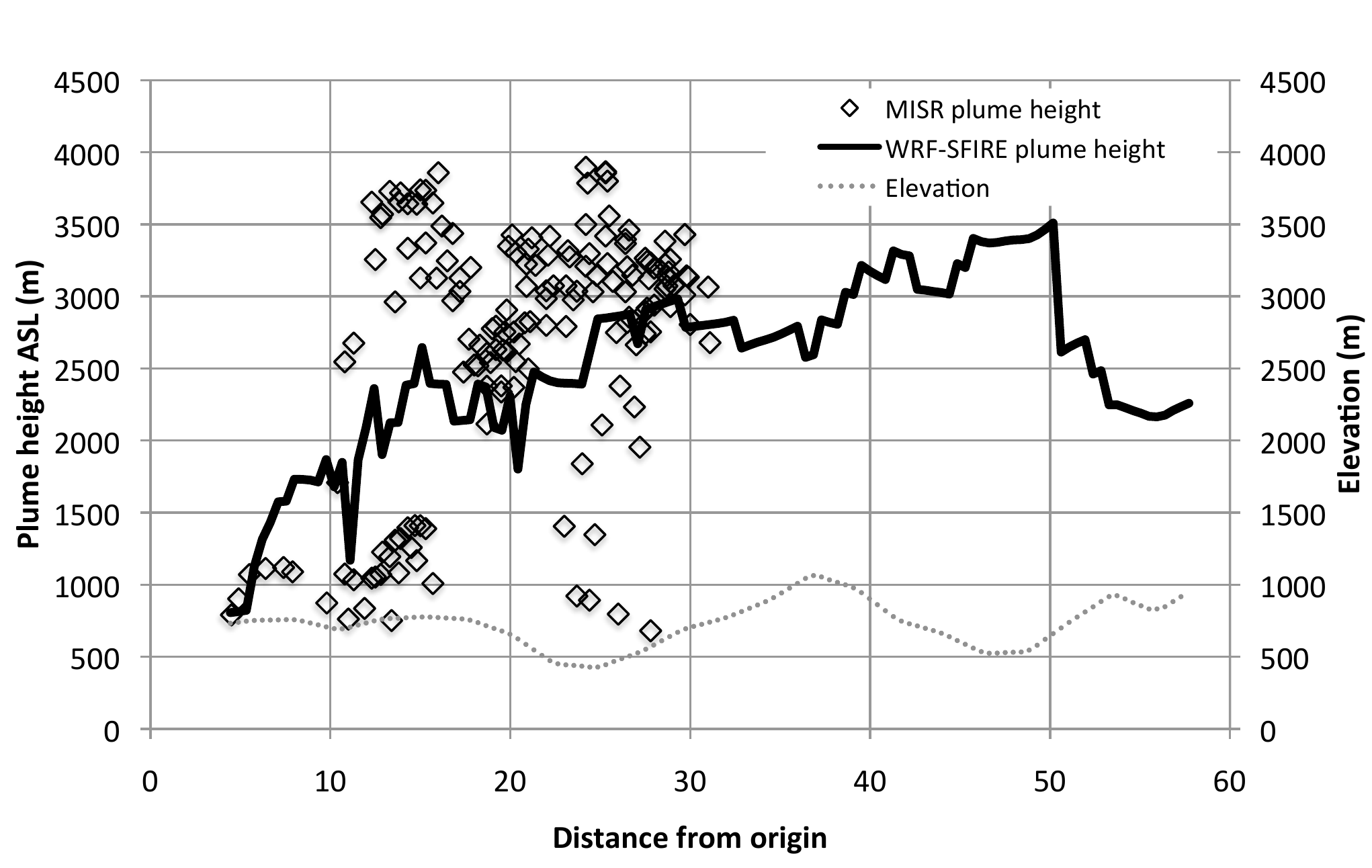}
 \end{center}
 \caption{Downwind plume height observed by MISR (open diamonds) and simulated by WRF-Sfire (solid line). 
Observed plume heights are based on MISR retrieval O67722-B52 on 2012-09-10 19:12 UTC, and
simulated plume heights from model output on 2012-09-10 19:15 UTC.   
Note elevation is given in m ASL (Above Sea Level).
\label{fig:Barker_Pl_Hts}}

\end{figure*}
 
In this case, instead of using the full chemical smoke representation, 
a passive tracer was used to visualize smoke rise and dispersion. 
The modelled plume was estimated applying a threshold tracer concentration of 0.5. 
The highest model level of a tracer concentration above this threshold was used to compute 
the WRF-Sfire with passive tracers modelled plume-top height.
One MISR overpass (067722-B52, date 2012-09-10) captured the Barker Canyon fire plume.
The MISR retrieval indicated a maximum plume-top height of 3896 m ASL (Above Mean Sea Level), 
a median plume height of 2008 m ASL, and a plume-top height standard deviation of 
852 m, for a smoke plume spreading over terrain heights between approximately 400 to 1000 m ASL.
Both model and MISR retrieval revealed that the plume-top heights fluctuated significantly
as the smoke-plume emissions advected downwind of the fire.
Therefore, for a better comparison between the simulated and MISR observed 
plume heights, both at their distance from the fire source are presented 
in Fig.~\ref{fig:Barker_Pl_Hts}. 
Scatter in the MISR plume-top heights is due to multiple data points at
the same distance from the fire origin.
The MISR data suggest that the Barker Canyon fire plume reached its maximum 
height of 3900 m ASL at 25 km downwind from its origin,  
while the WRF-Sfire plume reached its maximum height of 3500 m ASL 
approximately 50 km from its origin. 

From the standpoint of the air quality systems operating on grids of tens of 
kilometers, the differences between these two plume-top heights and their
distances from the fire source are relatively small.
There are multiple reasons for differences, but 
the primary reason for the difference in plume-top heights
may be that the coarse resolution of the WRF atmospheric grid did not represent 
adequately the impact of strong fire-induced heating (confined to the narrow fireline) 
on fire convection.
Dilution of fire-induced heat in WRF's surface layer, where plume width is smallest,
may be responsible for the lower than observed plume rise.
It is possible that the near-surface WRF grid cell size did not resolve adequately 
the buoyancy and pressure gradient forces associated with plume rise.
Despite this, the WRF-Sfire modelled maximum plume height based on an inert tracer field
matches the MISR observed heights well, with an underestimation of less than 10\%. 
 
\section{Summary and conclusions} \label{sec:Conclusions}
 
This study couples WRF-Sfire with WRF-Chem to form WRFSC.
As a coupled fire-atmosphere forecast system, WRF-Sfire makes it possible to 
resolve fire spread, heat release during flaming combustion, fire convection,
fire emissions, and plume rise.
As a coupled atmosphere-chemistry forecast system, WRF-Chem makes it possible
to model the chemical processes associated with smoke or pollutants emitted by
wildfires, downwind dispersion, deposition and associated chemistry.
As a combination of WRF-Sfire and WRF-Chem, WRFSC makes possible, at every WRF time step,
a completely integrated forecast of wildfire spread, pyro-plume rise, fire emissions,
and short- and long- range transport, dispersion, and chemistry of wildfire-caused pollution.
WRF-Sfire replaces WRF-Chem's reliance on a plume-rise module, 
assumptions about the size and heat release of the fire.  

WRFSC was evaluated based on comparisons between available observations and 
the results of two wildfire simulations, a 48h WRFSC simulation of the 
2007 Witch-Guejito Santa Ana fires and a 96h WRF-Sfire simulation with passive tracers of 
the 2012 Barker Canyon fire.

Particular matter PM2.5 concentrations were used to
visualize smoke emissions from the Witch-Guejito fires.
PM2.5 dispersion indicates the ability of 
WRFSC to capture observed complex fire progression and smoke-plume behaviour.
A comparison of PM2.5 concentrations to a corresponding MODIS satellite  
image suggested that WRFSC was able to simulate long-range smoke-plume dispersion 
associated with the growing fire perimeter for the Witch-Guejito fires.

Primary pollutants PM2.5 and NO observed by air quality stations
within the fire domain were used to evaluate modelled concentrations.
Although simulated results underestimated station observations, 
especially for NO, the differences between observed and 
WRFSC-forecasted PM2.5 and NO concentrations were not unacceptable compared 
to the large uncertainties in standard WRF-Chem forecasts 
influenced by wildfire \citep{pfisteretal:art:2011}.

A stringent test of WRFSC's capabilities is an ozone forecast.
WRFSC again underestimated station observations of ozone concentrations
in terms of magnitude and changes with time.
A secondary pollutant such as ozone is difficult to forecast,
and several reasons are offered in Section \ref{subsec:MOZART} to explain
the discrepancies between model results and observations.
However, as discussed by \citet[and references therein]{jaffe_wigder:art:2012},
fire-related ozone concentrations are generally very difficult to predict 
due to the high sensitivity to the retaliative concentrations of its 
precursors (NOx and NMOC), importance of the mixing due to the downwind meteorology, 
interactions between radiation and photochemistry, as well as the impact of both the 
meteorological conditions and fuel characteristics on the combustion effciency.      

One of the most important aspects of simulating wildfire plume transport is
determining plume-top heights at which fire emissions are injected into the atmosphere.
PM2.5 concentration results used to visualize 
smoke emissions from the Witch-Guejito fire were used to also
calculate plume-top heights and vertical profiles of smoke emissions.
The range of Witch-Guejito plume-top heights estimated from PM2.5 
matched the MISR plume retrieval data for two other fires burning at approximately the
same time and in the vicinity of the Witch-Guejito fires.
 
Based on available computing resources,
the 48h WRFSC Witch-Guejito simulation with MOZART chemistry took approximately 30h
and the 96h WRF-Sfire Barker Canyon simulation took approximately 13h.
The WRFSC Witch-Guejito simulation was considerably more expensive 
computationally than the Barker Canyon fire simulation using inert 
scalar tracers to represent smoke. 
Both simulations ran faster than real time, with the Barker Canyon model
set-up showing the promise of operational application.
Barker Canyon plume-top heights based on the simple passive tracer smoke representation
underestimated MISR observed plume-top heights by less than 10\%.
These results suggest that, in addition to the size and heat release of the 
fire perimeter and its impact on plume-rise as it propagates, 
WRF-Sfire can provide, at reasonable computational cost, based on passive smoke tracers,
the emissions profile needed by external chemical models.
In present pollution forecasting operations, an external chemical model, such 
as CMAQ (Community Multiscale Air Quality modeling system), could then be 
used to compute smoke chemistry and larger-scale transport.

There are many reasons for discrepancies between observations and model 
data, and it is outside the scope of this paper to determine exactly what they are.
Certainly many result immediately from the simplified WRF-Sfire and 
WRF-Chem setup designed as a first test of WRFSC.  

To isolate the impact of the simulated fire emissions on air quality,
the simulations did not use the chemical boundary conditions and 
standard idealized chemical profiles provided by WRF-Chem for atmospheric-chemistry
initialization.
All anthropogenic and biogenic emissions were omitted. 
Fuel moisture was kept constant throughout the Witch-Guejito simulation.
Both fire simulations were done without updating the model 
state with current meteorological observations during the run. 
Untested, emission fluxes were computed simply as the products of the 
fuel-combustion rates and fuel-specific emission factors.
Sfire model does not distinguish between flaming and smoldering. 
The effect of slower burning fuels on the emissions is that they emit smoke longer.
This simplified treatment of the link between the fuel consumption and fire emissions 
may not be adequate for fires with significant emissions during the smoldering stage. 
It is not known if three MODIS Land Cover Types are enough to represent the 
chemical emissions from wildfires spreading across thirteen possible fuel-bed 
categories, and WRF-Chem's global emission factors for certain fuel types
may differ from actual emission factors for those fuel types in certain geographic regions.
It is possible that the simulation period was too short for the chemical model to spin up. 
Simulated plume rise is sensitive to the atmospheric model cell size.  
When a horizontal grid as large as 500m is used in the fire domain,
the size of the near-surface grid volume may dilute the 
strength of fire heating input into the surface level of atmospheric model WRF 
enough to under-represent temperature and buoyancy forces in the plume.
Simulated plume-rise may be underestimated.
It is not known what horizontal grid resolution is required to reach a 
model plume rise insensitive to grid-size.

 
Most of these issues summarized above can possibly be mitigated with 
access to greater and more efficient computing resources.
This is only pilot study, using a very simple model setup,
designed to demonstrate the potential of WRF-Sfire and WRF-Chem as an integrated 
system for use by natural resource managers. 
What the study does establish is the increased level of detail provided by the system, 
such as locations of high reaction intensity, smoke emissions, and 
plume injection heights that can provide a more comprehensive understanding of the 
wildfire environment, wildfire behaviour, and downwind ramifications of 
wildland fire emissions on air quality.

Future quantitative research is warranted to prove the validity WRFSC, and
evaluation of the model by comparison to new and different data sets is necessary.
Observations that provide information for all components of 
the system --- local micrometeorology, fire spread, 
fire emissions, plume rise, and dispersion and chemistry --- are needed.
Coordinated field measurements of fire-spread data, fire-heat release, and
in-situ meteorological conditions are required to evaluate the fire spread component 
of the system. 
Radiosonde data are needed to provide information on the vertical structure of wind, 
moisture and temperature.
Airborne measurements in the smoke plume of the updraft velocities, 
temperature, and chemical composition, combined with estimates of local emission factors,
are needed to analyze fire-plume dynamics, dispersion, and chemistry.
A comprehensive dataset providing information on all these aspects is essential
to fully understand how the integrated system performs and what components
need to be improved. 
FireFlux \citep{clements_etal:art:2007} and RxCADRE \citep{ottmar:techreport:2013} 
experiments provide the most comprehensive datasets available for model evaluation 
purposes.  
Unfortunately, the former provides meteorological insight without
airborne smoke data, while the latter (fire L2f) is a great source of the smoke and fire
data but without in-plume meteorological measurements of the updraft
velocities and temperature.
Future field experiments are needed to provide comprehensive data sets 
for extensive evaluation and validation of WRF-Sfire, WRF-Chem, and WRFSC.
 
 
 

\bigskip\noindent ACKNOWLEDGMENTS.
This research was partially supported by the National Science Foundation (NSF) 
grants AGS-0835579 and DMS-1216481, and National Aeronautics and Space 
Administration (NASA) grants NNX12AQ85G and NNX13AH9G. 
This work partially utilized the Janus supercomputer, supported by the NSF grant CNS-0821794, 
the University of Colorado Boulder, University of Colorado Denver, and National Center 
for Atmospheric Research. 
\newpage
\bibliographystyle{jas99_m.bst}
\bibliography{rayleigh} 




\end{document}